\documentclass[sigconf,nonacm]{acmart}

\usepackage{bbm}

\usepackage{enumitem,kantlipsum}
\usepackage{subfig}
\usepackage{multirow}
\usepackage{arydshln}
\usepackage{natbib}
\usepackage{xcolor}
\usepackage[bookmarksnumbered,unicode]{hyperref} 

\AtBeginDocument{%
  \hypersetup{
    citecolor=Red,
    linkcolor=Blue,   
    urlcolor=Magenta}}
    
\AtBeginDocument{%
  \providecommand\BibTeX{{%
    \normalfont B\kern-0.5em{\scshape i\kern-0.25em b}\kern-0.8em\TeX}}}

\setcopyright{acmcopyright}
\copyrightyear{2018}
\acmYear{2018}
\acmDOI{XXXXXXX.XXXXXXX}

\acmConference[WWW]{ACM Web Conference}{May 13--17,
  2024}{Singapore}
\acmPrice{15.00}
\acmISBN{https://doi.org/10.1145/3543507.3583189}

\begin{document}

\title{BotSSCL: Social Bot Detection with Self-Supervised Contrastive Learning}


\author{Mohammad Majid Akhtar}
\email{majid.akhtar@unsw.edu.au}
\affiliation{%
  \institution{University of New South Wales}
  \country{Australia}
}
\author{Navid Shadman Bhuiyan}
\email{navid.bhuiyan@student.unsw.edu.au}
\affiliation{%
  \institution{University of New South Wales}
  \country{Australia}
}
\author{Rahat Masood}
\email{rahat.masood@unsw.edu.au}
\affiliation{%
  \institution{University of New South Wales}
  \country{Australia}
}
\author{Muhammad Ikram}
\email{muhammad.ikram@mq.edu.au}
\affiliation{%
  \institution{Macquarie University}
  \country{Australia}
}
\author{Salil S. Kanhere}
\email{salil.kanhere@unsw.edu.au}
\affiliation{%
  \institution{University of New South Wales}
  \country{Australia}
}
\renewcommand{\shortauthors}{Akhtar et al.}

\begin{abstract}

The detection of automated accounts, also known as ``social bots'', has been an increasingly important concern for online social networks (OSNs). While several methods have been proposed for detecting social bots, significant research gaps remain. First, current models exhibit limitations in detecting sophisticated bots that aim to mimic genuine OSN users. Second, these methods often rely on simplistic profile features, which are susceptible to manipulation.
In addition to their vulnerability to adversarial manipulations, these models lack generalizability, resulting in subpar performance when trained on one dataset and tested on another. 

To address these challenges, we propose a novel framework for social {\bf Bot} detection with {\bf S}elf-{\bf S}upervised {\bf C}ontrastive {\bf L}earning (BotSSCL). Our framework leverages contrastive learning to distinguish between 
social bots and humans in the embedding space to improve linear separability. 
The high-level representations derived by BotSSCL enhance its resilience to variations in data distribution and ensure generalizability. We evaluate BotSSCL's robustness against adversarial attempts to manipulate bot accounts to evade detection. Experiments on two datasets featuring sophisticated bots demonstrate that BotSSCL outperforms other supervised, unsupervised, and self-supervised baseline methods. 
We achieve $\approx6\%$ and $\approx8\%$ higher (F1) performance than SOTA on both datasets. In addition, BotSSCL also achieves 67\% F1 when trained on one dataset and tested with another, demonstrating its generalizability. 
Lastly, BotSSCL increases adversarial complexity and only allows 4\% success to the adversary in evading detection. 

\end{abstract}

\maketitle
\vspace{-0.2cm}
\section{Introduction}
OSNs have become one of the first contact points for content consumption, attracting billions of users. However, the open nature of OSNs also makes them vulnerable to orchestrated adversarial campaigns, frequently facilitated by automated accounts referred to as \textit{social bots}. These bots have played a significant role in spreading false information, particularly during the COVID-19 pandemic, leading to the dissemination of incorrect remedies, unproven practices, and various conspiracy theories. Consequently, trust in legitimate sources has been undermined~\cite{majidlcn}. Hence, there is a need for the early, effective, and efficient detection of social bots.

Numerous methods have been introduced to detect social bots~\cite{botometer,bothunter,digitaldna,satar,deeprobot}, spanning the gamut of supervised, unsupervised, and self-supervised approaches and demonstrating comparable performances on many datasets. Notably, to evade detection, bots often mimic genuine OSN users. These deceptive bots, termed as \textit{sophisticated bots}, have introduced three significant challenges.

First, a common approach employed by most of the aforementioned approaches involves extracting features from OSN accounts and mapping them into a feature space where social bots should ideally be distinguishable from human accounts. However, the behavior of sophisticated bots often leads to their closer proximity to human accounts in this feature space, which presents challenges related to linear separability ~\cite{Yang_Varol_Hui_Menczer_2020}. Real-world Twitter (rebranded as $\mathbb{X}$) datasets (such as Varol~\cite{varol2017online} and Gilani~\cite{gilani})
 that include these sophisticated bots showing high similarity between bots and humans when inspected under low-dimensional t-SNE plots as shown in the Appendix Figure~\ref{fig:problem}. Consequently, many existing works have achieved low accuracy in detecting bots in these two datasets~\cite{benfordlaw,tabassum2023many,guo2021social,ensemblebotometer,lobo,deeprobot}. 

Second, detecting sophisticated social bots within different datasets is challenging as bot profiles exhibit varied characteristics and behaviors according to their goals~\cite{akhtar2023false, rovito2022evolutionary}. Most existing models rely on supervised training, which often overfits on a specific bot dataset and makes the model less generalizable to detect bots that differ from the training data~\cite{lobo, Yang_Varol_Hui_Menczer_2020,rovito2022evolutionary,rauchfleisch2020false}. Generalizing these supervised models is further complicated by the arduous and error-prone task of data labeling. Third, it is common for adversaries to leverage AI tools to evade detection~\cite{cresci2020detecting,cresci2020decade}. An example could be a bot manipulating their profile to appear like a genuine OSN user~\cite{mou2020malicious} or using a Large Language Model (LLM) to post realistic content~\cite{ferrara2023social,yang2023anatomy}. Such adversarial attacks allow bots to evade detection from current models~\cite{lombardi2022ai,cresci2021coming}. 
Subsequently, novel approaches to address such problems are deemed pertinent.

This paper aims to detect sophisticated social bots with high accuracy. To this end, we propose BotSSCL (cf.  \S\ref{sec:botsscloverview}), a novel self-supervised technique grounded in contrastive learning. Our framework leverages contrastive learning, 
recognized for its capacity to address linear separability challenges~\cite{jaiswal2020survey}. While contrastive learning has previously demonstrated remarkable success in image-processing~\cite{simclr,jing2020self} and NLP domains~\cite{zhang2022contrastive,wu2020clear}, our research marks the pioneering application of contrastive learning to real-world tabular datasets for detecting sophisticated bots.
BotSSCL operates as a self-supervised learning system, effectively mitigating overfitting by eliminating the need for ground truth labels during training. 
 
\vspace{-0.08cm}
BotSSCL employs a contrastive-based framework, 
as it seeks to distinguish between bot and human samples. Our framework aims\vfill\break to organize similar data points in the embedding space by pulling together `anchor' points (that serve as reference points) and their corresponding `positive' samples belonging to the same class. Simultaneously, it works to push apart dissimilar points, represented by `negative' samples from different classes. 
Essentially, these anchor and positive samples, or different `views' of the same data point, share only the minimum information essential for the 
classification of sophisticated bots. To achieve this, we employ InfoNCE loss~\cite{simclr}, a contrastive loss function (cf. \S\ref{subsec:contrastiveloss}) specifically designed to maximize mutual information between anchors and positives, a crucial element in detecting sophisticated bots. 

\textbf{Contributions}: 
This framework introduces contrastive learning in the context of OSN bot detection and is both generic and robust when faced with adversarial attacks, providing a practical solution to the challenges presented by social bots in the domain of social media and online interactions.

So far, contrastive-based techniques are popular in the image and NLP domains~\cite{jing2020self,zhai2019s4l, simclr}. However, self-supervised contrastive learning (SSCL) is limited to \textit{tabular feature} datasets, which do not have the same structure as image and language data. 
Overall, this paper makes the following main contributions:
\begin{enumerate}[wide, labelwidth=1em, labelindent=0pt]
    \vspace{-0.7cm}
    \item {\bf Novel Framework Outperforming SOTA.} We propose our BoTSSCL, a novel framework for detecting sophisticated social bots. Diverging from conventional supervised detection methods, which frequently grapple with issues related to generalizability, our framework leverages a contrastive learning approach. We conduct an extensive evaluation and ablation study, showcasing the effects of various design choices and hyperparameters on two tabular datasets (Varol and Gilani). In comparison (cf. \S\ref{subsec:performance}), we outperformed five state-of-the-art (SOTA) on both datasets with ~8\% higher F1-score on the Gilani dataset (the more complex dataset). 
    \item {\bf Robust Generalization, Alleviating the Challenges of Data Labeling.} 
    Generalizability is achieved as BotSSCL follows the \textit{InfoMin} principle, which argues that maximization of mutual information is only helpful if the information is task-relevant~\cite{tian2020makes}. We empirically analyzed (cf. \S\ref{subsec:generalizability}) a reverse U-shaped curve (Figure~\ref{fig:corruption_rate}) that shows the relationship between mutual information and representation quality for detecting bots. The peak point in the curve denotes the sweet spot where the learning is optimal to detect bots. Learning beyond this point may include noise, which we avoid. BotSSCL achieves $\approx67\%$ F1-score when trained on the Varol dataset and tested on the Gilani dataset and vice-versa. Thus, we show that BotSSCL is dataset-agnostic and generalizable in detecting sophisticated bots across datasets.
    \item {\bf Robustness Against Adversarial Manipulations.} 
     Lastly, to account for potential evasion attempts, we consider adversaries as part of our threat model and explore their capabilities to manipulate the features employed in bot detection. Rather than relying solely on raw profile features, BotSSCL involves the construction of a normalized user representation achieved through the concatenation and transformation of diverse feature sets via linear transformations. This process results in mapping input features to arbitrary output dimensions, creating an additional layer of complexity for potential adversaries seeking to make substantial modifications to their profiles. Our findings reveal (cf \S\ref{subsec:adversarial}) the intricacy and difficulty faced by adversaries, even those with modest resources, achieved only 4\% success rate when attempting adversarial attacks. 
\end{enumerate}

\section{Methodology of BotSSCL}
\label{sec:botsscloverview}
{\bf Problem Statement.} Given training data that consists of OSN accounts ${U}$ and ${T}$ posts (or tweets), where ${U}$ is the number of accounts denoted as $U=\{u_{1},u_{2},u_{3}...,u_{u}\}$  and \textit{T} is the number of posts (or tweet) data denoted as $T_{u}=\{t_{1},t_{2},t_{3},...,t_{u}\}$, we aim to detect bot accounts in the testing data. 
The output of BotSSCL is a set of binary labels
indicating whether each account is a bot or not,
i.e., $y(U) \in \{0,1\}$, where $y(U) = 1$ indicates that user (interchangeably referred to account) \textit{$u_i$} is a bot. 

\begin{figure*}[!ht]
  \centering
  \includegraphics[width=\linewidth]{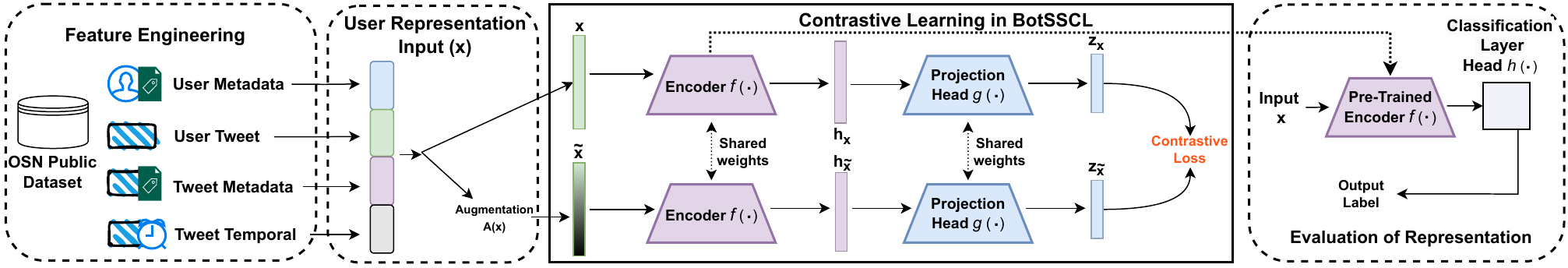}
  \vspace{-0.40cm}
  \caption{Overview of BotSSCL framework. The first and second block (from left to right) represents the feature engineering and user representation process. The third block denotes contrastive learning. In the fourth block, we evaluate the representations. }
  \label{fig:botsscl}
  \vspace{-0.4cm}
\end{figure*}

{\bf BotSSCL Framework.} To address the aforementioned problem, BotSSCL is expected to have a set of functional requirements and attributes, explicitly about generalizability, self-supervised nature, and resilience against adversary manipulations. Figure~\ref{fig:botsscl} depicts the methodology, encompassing our augmentation technique, various contrastive loss variants, and the metrics employed to assess BotSSCL in attaining these functional requirements. 

\subsection{Feature Engineering}
\label{subsec:feature_engineering}
Various models have been developed to differentiate between bot and human accounts on OSNs such as Twitter. These models utilize various combinations of features that can be extracted from Twitter profiles or accounts. 
Building upon previous work~\cite{beskow2019its}, we categorize feature engineering into three distinct tiers. Table~\ref {tab:features_used} represents different data collection tier levels and the associated derived features. As shown in the table, Tier 0 includes the collection of only metadata about the OSN accounts and requires the least time and resources. In contrast, Tier 2 refers to the additional API request that extracts the network-related information, such as the following relationship. This additional data brings structural data but also increases the dependency on API calls;
particularly with the recent introduction of a limit at which the Twitter API can be used to initiate queries. {Our feature engineering method falls under the Tier-1 category,
where we collected four types of features from social accounts:

\begin{enumerate}[wide, labelwidth=1em, labelindent=0pt]
    \item \textbf{User Metadata:} includes Twitter user profile features to capture the unique statistics of each account, such as the number of followers of an account and the number of other accounts it follows. We extracted 33 user metadata features, denoted as $f$. 
    \item \textbf{User Tweets:} involves the conversion of the user's textual tweets into embeddings. 
    We only extracted 200 recent tweets of users considering Twitter API limits and are denoted as $T_{u}$.
    \item \textbf{Tweet Metadata:} includes features such as the number of likes and retweets each tweet receives. We derived 29 metadata features, denoted as $t_{m}$.
    \item \textbf{Tweet Temporal:} includes features, such as the time between tweets, to capture an account's temporal relationships and periodic behavior. We used seven temporal features, denoted as $t_{t}$.  
\end{enumerate}

Table~\ref{tab:features_elaborated} in Appendix ~\ref{appendix:features_used} provides a comprehensive list of the feature names employed by BotSSCL at Tier 1.

\vspace{-0.2em}
\begin{table}[h]
\vspace{-0.20cm}
   \caption{Features used in Twitter Bot Detection.} 
   \vspace{-0.20cm}
\label{tab:features_used}
\scalebox{0.7}{
  \begin{tabular}{r :c c c c c}
   \toprule
  \textbf{Features}& \textbf{Tier 0}& \textbf{Tier 1}& \textbf{Tier 2}\\
   \hline
   User Metadata & \checkmark & \checkmark & \checkmark \\
   User Tweet & $\times$ & \checkmark & \checkmark \\
   Tweet Metadata & $\times$ & \checkmark & $\checkmark$ \\
   Tweet Temporal & $\times$ & \checkmark & $\checkmark$ \\
   Network Information & $\times$ & $\times$ & \checkmark \\
   \bottomrule
  \end{tabular}
  }
\end{table}

\vspace{-1em}
\subsection{Twitter User Representation}
\label{subsec:twitteruserrep}
Once the features at Tier 1 are extracted, we generate a normalized representation vector that fully represents a specific Twitter user. 
The benefit of using a user representation vector is twofold. Firstly, it effectively combines various user information modalities, including user metadata, user tweet content, tweet metadata, and tweet temporal features--to enhance bot detection accuracy and fortify resilience against adversarial attacks, as illustrated in Figure~\ref{fig:twitter_user_representation}.
Second, we perform linear transformations on the extracted features 
to make our bot detection framework robust against feature manipulation attacks.
Our suggested Twitter user representation comprises four feature categories, each making up a quarter of the user representation. This means that if the \textit{`D'} input feature dimension (or information density) is 16, the Twitter user representation dimension would be (16$\times$4=64).
 An adversary may still attempt to manipulate their profile, resulting in different user representations. We further investigate changes in the representation in (cf.~\S\ref{subsec:adversarial}). We define representations below:

\subsubsection{User Metadata Feature}
In total, we have $|U|$ users and $|f|$ user's metadata features $ \forall u \in U$. We first apply \textit{z-score normalization}~\cite{pedregosa2011scikit} to adjust the values of features to a common scale. For example, \textit{number of followers} is a numerical feature that can have a higher value, whereas a feature, \textit{number of digits in the username} will have a lower value. 
Next, the user metadata feature is mapped into an input vector $m \in \mathbb{R}^{|U| \times |f|} $. Then, the vector $m$ is projected into $r_{um} \in \mathbb{R}^{|U| \times D}$ by a trainable linear layer, as in Eq.~\ref{eq:1}.
\begin{equation}
r_{um} = (m W_{1}^T + b_{1})
\label{eq:1}
\end{equation}
where $W_{1} \in \mathbb{R}^{D \times |f|}$ represents weight matrix, $b_{1} \in \mathbb{R}^{D}$ represents bias matrix, and $r_{um} \in \mathbb{R}^{|U| \times D}$ represents user metadata representation. \textit{D} is the input feature dimension, i.e., 33 user metadata features for an account are transformed into \textit{D} input dimension. 

\subsubsection{User Tweets Feature}
For every user $u \in U$, we have $|T_{u}|$ number of user's Tweets where we first encode each Tweet $t_{i} \in T_{u}$ using BERT pre-trained model~\cite{reimers-2019-sentence-bert} to get the fixed sized of 768 dimension Tweet's embedding vector $t_{e} \in \mathbb{R}^{768}$. Next, we average all Tweet embeddings for all the users, as follows: 
\begin{equation}
t_{e}^{u} =  \frac{1}{|T_{u}|} \sum\limits_{i=1}^{|T_{u}|} BERT(t_{i}^{u}), \forall u \in U
\label{eq:2}
\end{equation}
The Tweets embedding vector is projected into $r_{ut} \in \mathbb{R}^{|U| \times D}$ by a trainable linear layer, as shown in Eq.~\ref{eq:3}.
\begin{equation}
r_{ut} = (t_{e} W_{2}^T + b_{2})
\label{eq:3}
\end{equation}
where $W_{2} \in \mathbb{R}^{D \times 768}$  is a weight matrix, $b_{2} \in \mathbb{R}^{D}$ is a bias matrix, and $r_{ut} \in \mathbb{R}^{|U| \times D}$ is user's Tweet representation, whereas \textit{D} represents the feature dimension. 
\vspace{-0.2cm}
\subsubsection{Tweet Metadata Feature}
We have $|t_{m}|$ metadata features of user's tweets $ \forall u \in U$. Similar to prior steps, we first apply \textit{z-score normalization} to normalize the features. Then, we project the vector $t_{m}\in \mathbb{R}^{|U| \times |t_{m}|}$ into $r_{um} \in \mathbb{R}^{|U| \times D}$ by a trainable linear layer, as,
\begin{equation}
r_{tm} = (t_{m} W_{3}^T + b_{3})
\label{eq:4}
\end{equation}
where $W_{3} \in \mathbb{R}^{D \times |t_{m}|}$, $b_{3} \in \mathbb{R}^{D}$ and $r_{um} \in \mathbb{R}^{|U| \times D}$. 
\vspace{-0.1cm}
\subsubsection{Tweet Temporal Feature}
In total, we have $|t_{t}|$ number of temporal features of the user's tweets for all $ \forall u \in U$. Again, we first apply \textit{z-score normalization} to scale features. Then, $t_{t}\in \mathbb{R}^{U \times |t_{t}|}$ is projected into $r_{tt} \in \mathbb{R}^{|U| \times D}$ via a trainable linear layer, as in Eq.~\ref{eq:5}.
\begin{equation}
r_{tt} = (t_{t} W_{4}^T + b_{4})
\label{eq:5}
\end{equation}
where $W_{4} \in \mathbb{R}^{D \times |t_{t}|}$, $b_{4} \in \mathbb{R}^{D}$ and $r_{tt} \in \mathbb{R}^{|U| \times D}$. Here also, \textit{D} represents the feature dimension.

Finally, we concatenate all four transformed \textit{D} dimension vectors to form a Twitter user representation (of \textit{D$\times$4} dimensions) as: 
\begin{equation}
t_{rep} = [r_{um};r_{ut};r_{tm};r_{tt}]
\label{eq:6}
\end{equation}

where $r_{um}$~is \textit{user metadata} representation, $r_{ut}$~is \textit{user tweet} representation, $r_{tm}$~is \textit{tweet metadata} representation and $r_{tt}$~is \textit{tweet temporal} representation. Hence, we redefine the objective of BotSSCL as follows: Given Twitter user representation of training data $t_{train}^{rep} = [t_{train}^{(1)},t_{train}^{(2)},...,t_{train}^{(U_{train})}]$, our goal is to detect bot, $y(U)$ in testing data $t_{test}^{rep} = [t_{test}^{(1)},t_{test}^{(2)},...,t_{test}^{(U_{test})}]$. Next, we use these user representations in contrastive learning.
\vspace{-0.43cm}
\begin{figure}[h]
  \centering
  \includegraphics[width=0.75\linewidth, height=5.8cm]{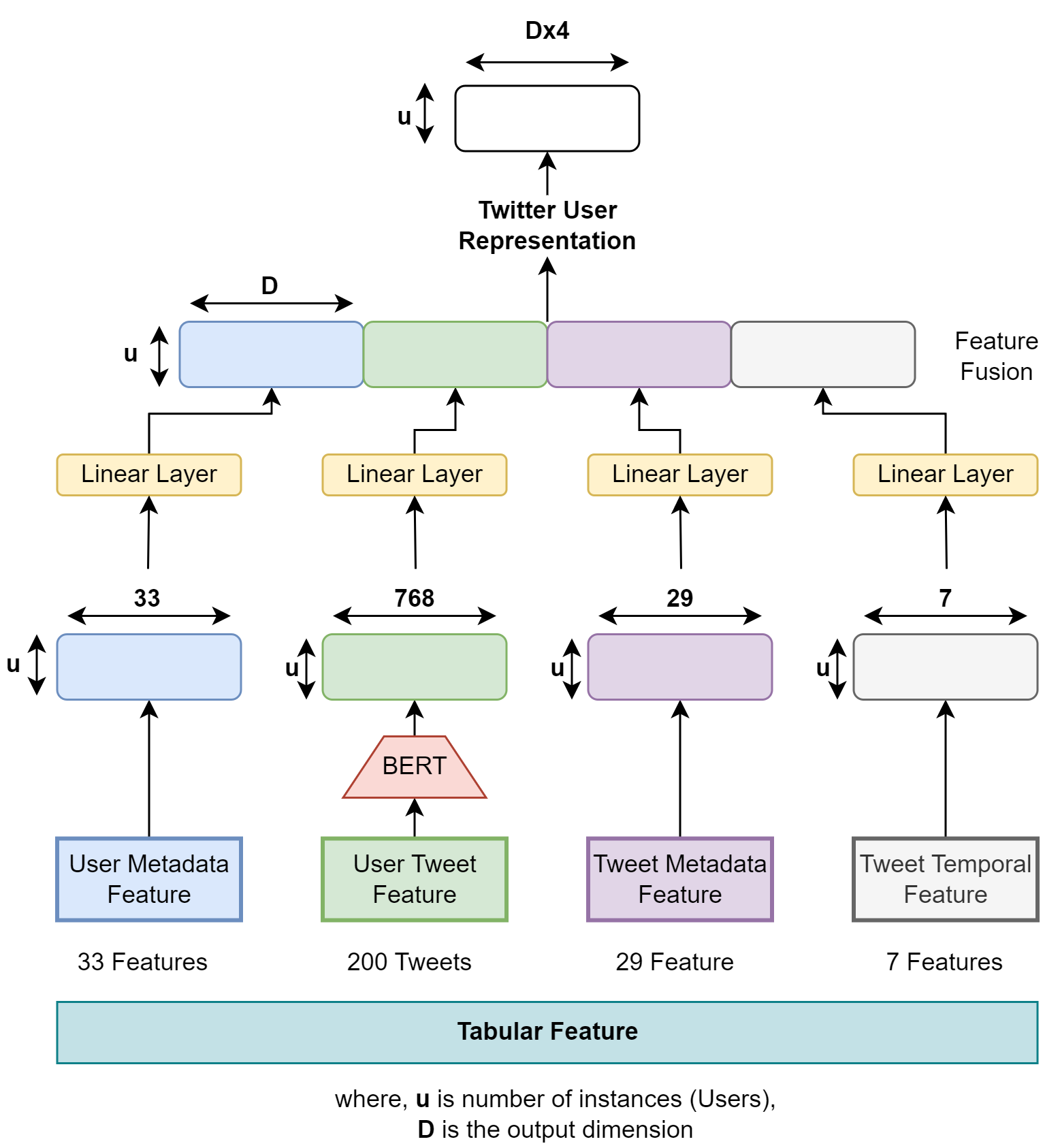}
  \vspace{-0.35cm}
  \caption{Twitter User Representation}
  \label{fig:twitter_user_representation}
\end{figure}

\subsection{Contrastive Learning in BotSSCL}
\label{subsection:workingbotsscl}
Our contrastive self-supervised learning framework consists of a twin encoder network, transforming the Twitter user representation into informative embeddings. The projection head, denoted as $\textit{g}\ (\cdot)$, is designed to follow the encoder $\textit{f}\ (\cdot)$. The encoder processes the output generated by the encoder network and projects it into a feature embedding space (unit hypersphere). The encoder and the projection head are multi-layer perceptrons (MLP), where the encoder consists of two layers, and the projection head contains one layer. Each layer, excluding the first one, is equipped with \textit{D} neurons. Specifically, the first layer has \textit{D$\times$4} neurons, corresponding to the dimension size of the Twitter user representation.

For contrastive training, every sample $x_{i}$ is treated as an anchor, representing a bot or human account serving as a reference point within a given mini-batch. These anchor samples are subjected to a data augmentation technique, $a(x_{i})$, to generate an augmented version denoted as ($\tilde{x_{i}}$), which is treated as the positive. Notably, we have used the terms `positive' and `views' interchangeably, both referring to the augmented version of the anchor point. These samples are subsequently passed through two encoder networks, which share weights and biases, as well as projection heads. This process results in obtaining embeddings $z_{x_{i}}$ and $z_{\tilde{x_{i}}}$, as depicted in Figure~\ref{fig:botsscl}. The training involves a contrastive loss, specifically the InfoNCE error~\cite{simclr}, designed to push dissimilar data apart while pulling similar data within the mini-batch closer.

The advantage of utilizing the contrastive loss lies in its ability to learn representations essential for a particular downstream task (\textit{T}). In our case, the task involves predicting labels $y(U)$ for accounts. In practice, optimal views (containing only task-relevant information) are found at a specific `sweet spot,' maximizing \textit{I}(x; $y(U)$) and \textit{I}($\tilde{x_{i}}$; $y(U)$), while minimizing \textit{I}(x; $\tilde{x_{i}}$)-- this aligns with the \textit{InfoMin} principle~\cite{tian2020makes}. In essence, contrastive learning aims to develop a parametric function that can effectively discriminate between bots and humans. This is achieved through the fine-tuning of the parameters within functions \textit{f} and \textit{g} using stochastic gradient descent (SGD). Please refer to Appendix~\ref{appendix:experimentsetup} for detailed implementation specifications. The subsequent sub-sections will define the augmentation techniques used in this work.

\subsubsection{Data Augmentations}
In contrastive learning, the augmentation process involves three key components. First, the original data serves as an anchor. Second, the original data is subject to augmentation, generating an alternative view marked as positive. Third, the remaining images within the batch are treated as negatives. 

The most commonly used augmentations in literature have been successfully applied to image-based scenarios as augmentations preserve their inter-pixel relationship~\cite{simclr,jaiswal2020survey}.
On the other hand, tabular data feature augmentation is an active area of research in NLP and presents unique challenges compared to image augmentation. Evaluating the quality of augmentation requires subjective assessments, which can be challenging. Nonetheless, we selected three augmentation methods suitable for our work. Since our work is based on tabular features, we adopted one augmentation technique from~\cite{bahri2021scarf} and present two new augmentation techniques for the NLP community in this work. Note that the contrastive objective depends on the correct type of augmentation technique to generate positive; therefore, we tested the three augmentations in our BotSSCL framework. The augmentations are the following:

{\bf Augmentation 1 -- Random Feature Corruption:} 
As the name suggests, for each sample (acting as an anchor) in the mini-batch (a subset of training data in each iteration), we generate a corrupted version to use as the augmented (positive) sample. For this, we draw some random features of the anchor and replace each feature with that of another randomly selected sample from the dataset.  Figure~\ref{fig:augmentations} illustrates the augmentation 1 method. Our motivation to use this method is derived from Bahri et al.~\cite{bahri2021scarf}, whose research demonstrated that feature corruption preserves sufficient mutual information between views.

\textbf{Augmentation 2 -- MICE Imputation:}
In this data augmentation technique, we leverage the use of 
multivariate
imputation by chained equations (MICE)~\cite{van2011mice}. We first take the anchor sample and deliberately put NaN at 30\% of the places 
to preserve mutual information between the anchor and the positive data sample. We then used the MICE algorithm~\cite{van2011mice} that imputes the NaN values created in the anchor data samples, as shown in Figure~\ref{fig:augmentations}.
Note that this augmentation method may take a long computation time in practice as the MICE algorithm for large dimensions can be costly, so we kept replacement as 30\% 
The resultant is marked as positive, closely resembling the anchor. This technique is somewhat similar to missing feature corruption~\cite{bahri2021scarf}.

{\textbf{Augmentation 3 -- Linear Transformation:}}
In this augmentation type, we take dimension $N$ anchor data as input and pass it through a fully connected linear layer to output a vector of the same $N$ dimension. The benefit of this transformation is that it preserves the underlying linear structure of the vectors, which we consider as another view of the anchor. Thus, the resultant vector is marked as positive, as illustrated in 
Figure~\ref{fig:augmentations}. In terms of computation, this augmentation is faster than the augmentation~2.

\begin{figure*}[!htb]
    \centering
    \subfloat[\centering]{{\includegraphics[width=0.6\linewidth]{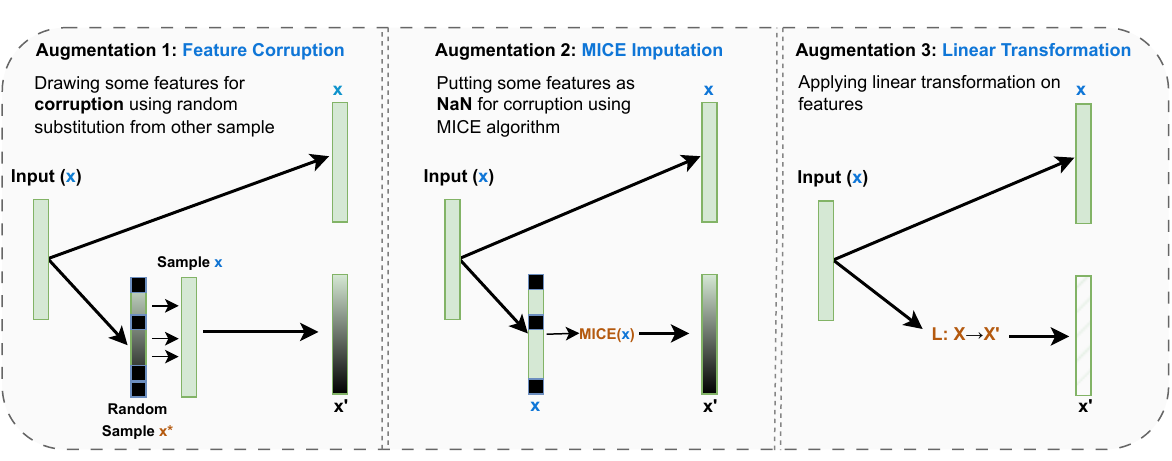} }%
    \label{fig:augmentations}}
    \subfloat[\centering]{{\includegraphics[width=0.3539\linewidth]{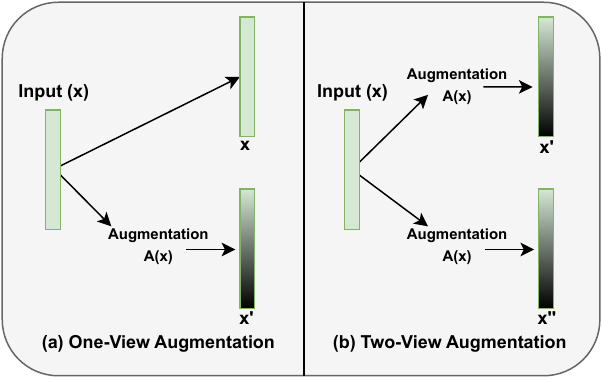} }
    \label{fig:one_to_two_augmentation}}
    \vspace{-0.3cm}
    \caption{a) Different Augmentations Types, b) One or Two-view level augmentation}%
    \label{fig:gilanirec}%
    \vspace{-0.4cm}
\end{figure*}

\subsubsection{Augmentation Settings:}
In contrastive learning, using positive and negative pairs encourages the model to learn discriminative features. As previously mentioned, the augmentation creates positives (as other views of the original data). This augmentation can be applied in two settings as follows:

\textbf{Two-view Augmentation:}
Typically, two-view augmentation is a more common approach in contrastive learning. The original sample (anchor, denoted as $x_i$) is passed through the augmentation technique twice to generate two views of the original data as positive samples ($\tilde{x_{i}}$ and $\tilde{\tilde{x_{i}}}$). Contrastive training then utilizes these two pairs. Figure~\ref{fig:one_to_two_augmentation}
shows the approach. 

\textbf{One-view Augmentation:}
In the second type, every sample in the dataset yields a single positive ($\tilde{x_{i}}$) using augmentation. 
The rationale behind using a one-view (positive) sample strategy alongside an anchor is to maintain a certain level of similarity compared to employing a two-view augmentation approach. Using two-view augmentation of the anchor has a high chance of reducing shared mutual information between two positive views. Hence, for our work, we used a one-view augmentation setting. 

\subsection{Contrastive Loss}
\label{subsec:contrastiveloss}
The contrastive loss encourages the model to bring positive samples close in the embedding space while pushing negative samples apart. We consider the following two variants of contrastive loss:

\subsubsection{Self-supervised contrastive loss}

In a mini-batch of size N, let $i \in \textit{I} = \{1,2,3....,N\} $ be the index of a sample and let $a(i)$ be the index of its augmented sample originating from the same source sample \textit{i}. The loss is defined as in \textit{self-supervised} contrastive learning (InfoNCE loss~\cite{simclr}).
\begin{equation}
    \mathcal{L}^{self} = {-}\sum\limits_{i \in I}\log {\frac{\exp {(sim(z_{i},z_{a(i)})/\tau)}}{ \sum\limits_{k \in A(i)}\exp {(sim(z_{i},z_{k})/\tau)}}}
\label{eq:7}
\end{equation}

Here, $z_{i}$ = $Proj(Enc(x_{i}))$, $z_{a(i)}$ = $Proj(Enc(\tilde{x_{i}}))$, where $\tilde{x_{i}}$ is the augmented sample from $x_{i}$, \textit{sim} is a similarity function, $\tau$ is a scaler temperature parameter, \textit{k} is the index of negative, and $A(i) = I \textbackslash \{i\}$. The index \textit{i} is the anchor, index \textit{a(i)} is the positive, and the rest of \textit{$\{A(i) \textbackslash a(i)\}$} indices are negatives. Therefore, for each sample \textit{i} there are $2(N-2)$ negatives. Thus, the denominator has $2(N-1)$ terms (all negatives and one positive)~\cite{khosla2020supervised}.

\subsubsection{Supervised contrastive loss}

Since negatives could also include samples that belong to the same class, in \textit{supervised} contrastive learning, the loss tries to bring all positive samples closer using the same class label information. The modified loss is as in~\cite{khosla2020supervised}.
\begin{equation}
    \mathcal{L}^{sup} = {-}\sum\limits_{i \in I}\frac{-1}{|Pos_{(i)}|}\sum\limits_{p \in Pos_{(i)}}\log {\frac{\exp {(sim(z_{i},z_{p})/\tau)}}{ \sum\limits_{k \in A(i)}\exp {(sim(z_{i},z_{k})/\tau)}}}
\label{eq:9}
\end{equation}
Here, $Pos_{i} = \{p \in A(i), where \ \tilde{y}_{p} = \tilde{y}_{i}\}$ is the set of all positives indices in the mini-batch for sample \textit{i} and $|Pos_{(i)}|$ is its cardinality. 

However, this also leads to another choice for the loss function to eliminate all positive terms from negative in the denominator with respect to sample \textit{i} to contrast with only negatives as $Neg(i) = \{k \in A(i), where \ \tilde{y}_{k} \neq \tilde{y}_{i}\}$, i.e., $2(N-|Pos_{(i)}|)$ terms. Thus, the new \textit{modified} loss becomes.
\begin{equation}
    \mathcal{L}^{sup}_{mod} = {-}\sum\limits_{i \in I}\frac{-1}{|Pos_{(i)}|}\sum\limits_{p \in Pos_{(i)}}\log {\frac{\exp {(sim(z_{i},z_{p})/\tau)}}{ \sum\limits_{k \in Neg(i)}\exp {(sim(z_{i},z_{k})/\tau)}}}
    \label{eq:10}
\end{equation}

\subsection{Evaluation Metrics}
Generally, the contrastive loss itself is an evaluation metric. A lower contrastive loss indicates better model performance. On the other hand, we can also assess the quality of the learned representations by visualizing them in a lower-dimensional space using techniques such as t-SNE. However, in most common works, learned representations are subjected to linear evaluation for some downstream tasks (in this case, bot classification). Typically, this involves training a simple classification head $\textit{h}\ (\cdot)$, such as logistic regression or a linear Support Vector classifier (SVC), on top of the pre-trained encoder $\textit{f}\ (\cdot)$. We optimize the cross-entropy loss and fine-tune the parameters of $\textit{f}\ (\cdot)$ and $\textit{h}\ (\cdot)$. The benefit is having a simple model with fewer learnable parameters than the complex neural network used for representation learning. Additionally, evaluation provides a comprehensive understanding of the model's performance and generalization capabilities. Therefore, we depend on logistic regression for the evaluation in this work. 
Furthermore, we use precision (Prec), recall (Rec), and F1-score (F1) over the test dataset to evaluate the performance of baselines and our framework. We set the $`class\_weight'$ parameter of logistic regression to `balanced'. Note that when the dataset is imbalanced, these metrics are appropriate~\cite{lones2021avoid, haixiang2017learning}.

\vspace{-0.1cm}
\section{Evaluation and Results}
\label{sec:evaluation}
In this section, we conduct experiments to answer the following research questions: 
\begin{itemize}[wide, labelwidth=!, labelindent=0pt]
    \item \textbf{RQ1 (Performance and Ablation):} Does BotSSCL outperform the baseline methods in the F1-score, and how do various design choices and hyperparameters contribute to the performance? 
    \item \textbf{RQ2 (Generalizability):} How does BotSSCL perform on the data that the model has not seen during the model training?
    \item \textbf{RQ3 (Adversarial Robustness):} Is BotSSCL robust to adversarial attacks? Can an adversary easily evade detection?
\end{itemize}

\vspace{-0.1cm}
\subsection{Datasets}

As already mentioned, deceptive bots try to mimic OSN humans. These bots tend to be projected closer to human accounts, posing challenges for linear separability~\cite{Yang_Varol_Hui_Menczer_2020}. Datasets such as the Varol and Gilani datasets reveal a striking resemblance between sophisticated bots and humans when examined through low-dimensional t-SNE plots, as illustrated in Appendix~\ref{appendix:visualization}, Figure~\ref{fig:problem}. Consequently, numerous existing studies have encountered difficulties achieving good performance on these two datasets~\cite{benfordlaw, tabassum2023many, guo2021social, varol2017online, ensemblebotometer, lobo,deeprobot}. Hence, we selected these two publically available datasets, Varol Dataset (referred to as D1) and Gilani Dataset (referred to as D2). 

We collected the datasets (D1 and D2) from Botometer repository~\cite{botometerrepository} that includes account ID's and labels (bot or human). We extracted the labels (bot or human) for each account in the dataset. However, we also required 200 Tweets and user profile metadata for every account not present in the D1 and D2 datasets. Therefore, we enhanced it by crawling more attributes directly from Twitter
for the accounts given in D1 on 5th March 2023 and D2 dataset on 7th March 2023. However, while doing so, we observed fewer accounts than the original number of accounts in D1 and D2. The reason is that a few accounts were inactive and thus not accessible. Also, for the Gilani dataset, 18 accounts were doubly labeled (to both classes) and thus removed to reduce ambiguity. Table~\ref{tab:dataset} summarises the details of the D1 and D2, where 4,389 accounts were extracted. Data and Code will be made available upon request.

\begin{table}[h]
  \centering
  \vspace{-0.20cm}
   \caption{Details of the publicly available annotated datasets.} 
   \vspace{-0.40cm}
\label{tab:dataset}
\scalebox{0.75}{
  \begin{tabular}{c c c c c c c}
   \toprule
   \multicolumn{2}{c}{\multirow{2}{*}{\textbf{Dataset Name}}}&\multirow{2}{*}{\textbf{Total Accounts}}&\multicolumn{2}{c}{\textbf{Train (80\%)}}&\multicolumn{2}{c}{\textbf{Test (20\%)}}\\ \cline{4-7}
  & & &\textbf{Human} & \textbf{Bot} &\textbf{Human} &\textbf{Bot}\\
   \hline
   \multicolumn{2}{c}{\textbf{Varol Dataset} (D1)~\cite{varol2017online}} & 2,074 (out of 2,573)  & 1,132 & 527 & 283 & 132\\
   \multicolumn{2}{c}{\textbf{Gilani Dataset} (D2)~\cite{gilani}}& 2,315 (out of 2,652) & 1,072 & 780 & 268 & 195
   \\ 
   \bottomrule
  \end{tabular}}
  \vspace{-0.40cm}
\end{table}

\subsection{Baselines}
We compare the performance of our proposed framework with five popular bot detection frameworks, ranging from supervised, unsupervised, and self-supervised models that cover diverse features from the three feature tiers. 

\textbf{Botometer (Tier 2)} is a supervised model that uses a random forest classifier to output a bot score using more than 1200 features~\cite{botometer}. The data collection requirement for feature engineering for this model falls under Tier 2 from Table~\ref{tab:features_used}. Using the Botometer API service with the Twitter account IDs, we retrieved scores and processed them to classify whether the account was a bot. Typically, 0.5 is considered a common threshold used in many prior works where above 0.5 is considered a bot, and a score underneath is considered a human~\cite{vosoughi2018spread,shao2018spread,shu2020fakenewsnet,wright2018don,majidlcn}. We also tested the dataset with different bot thresholds, i.e., 0.5, 0.6, and 0.7 (as shown in Table~\ref{tab:performance}). We found that performance (F1) decreases with an increasing threshold as it filters only a few accounts as bots compared to ground truth. 

\textbf{BotHunter (Tier 0)} is another random forest classifier (a supervised method) that utilizes user metadata features to analyze differences between bot and human accounts~\cite{bothunter}. Apart from user metadata features, it also derives a few other useful features, such as the average Tweet per day and user name entropy.

\textbf{DigitalDNA (Tier 1)} is an unsupervised method that mainly focuses on detecting coordinating bots based on Tweet posting~\cite{digitaldna}. The approach draws inspiration from DNA to create a unique fingerprinting and has two sub-models, B3 and B6. For instance, B3 represents a digital DNA sequence with only three alphabets, i.e., A to any form of a Tweet, character T to retweet, and C to reply. Similarly, B6 uses six characters for a digital DNA sequence, such as U for Tweets containing URLs.

\textbf{DeeProBot (Tier 0)} is a supervised hybrid deep neural network that uses user metadata features from Twitter user objects~\cite{deeprobot}. The method uses LSTM to process text from the user profile's description and joins it with user metadata features.  

\textbf{SATAR (Tier 2)} is a self-supervised model that generates labels by using all five categories of features, with a particular emphasis on network information features~\cite{satar}. It assumes bot behavior can be derived from its interactions with other accounts.

\subsection{Performance of BotSSCL (RQ1)}
\label{subsec:performance}
In Table~\ref{tab:performance}, we show the bot detection performance in terms of precision, recall, and F1-score of our BotSSCL and the baseline methods on the Varol and Gilani datasets. 
We tested our BotSSCL with Tier 0 (using only the user metadata feature) and Tier 1 (using user metadata, Tweet text, Tweet metadata, and Tweet temporal features). The results show that BotSSCL outperforms the baselines in both datasets. BotSSCL has $\approx8\%$ higher F1-score than the Botometer baseline on the Gilani dataset. Table~\ref{tab:performance} also shows that BotSSCL achieves the highest performance among all baselines, even with Tier 0 data collection on the Gilani dataset. However, the Varol Dataset requires Tier 1 data collection to perform best and achieves $\approx6\%$ higher F1-score than baselines, as shown in Table~\ref{tab:performance}. It is worth mentioning that Tier 1 features are still less resource-intensive (see \S~\ref{subsec:feature_engineering}) as we do not rely on network features, which are computationally
expensive for data collection~\cite{bothunter}. 
Therefore, our model shows effectiveness even with a dataset involving sophisticated bots replicating human behavior scenarios, which are highly important for OSN providers to detect. Note that the Gilani dataset has more homogeneous accounts than Varol and is difficult to classify using baselines, as shown in Appendix~\ref{appendix:visualization}, Figure~\ref{fig:problem}. 

Moreover, Table~\ref{tab:all_evalutation_varol} and Table~\ref{tab:all_evalutation_gilani}  in Appendix~\ref{appendix:performance} show all evaluation results with different values of \textit{D} input feature dimension and output embedding dimension projected by projection head in our framework. To make it easy to understand, (16$\times$16) denotes a test case (test ID) with \textit{D}=16 input dimensions (64 dimensions of Twitter user representation) and 16 output embedding dimensions. We observe that the Varol dataset achieves 80\% (F1) with test ID (16$\times$64).
In contrast, the Gilani dataset achieves 79\% (F1) with test ID (64$\times$64) as depicted in Table~\ref{tab:performance}.
It indicates that the two datasets perform best with different settings. 
However, to achieve uniformity and generalizability in bot detection, we keep the input dimension as D=64 and output embedding to 64 as the default setting for both datasets for all the tests unless specified otherwise. With (64$\times$64) dimensions, BotSSCL still achieves 77\% (F1) on the Varol dataset and achieves 79\% (F1) on the Gilani dataset (harder to classify dataset), showing the highest performance among other baselines.
BotSSCL performs better than supervised (Botometer, BotHunter, and DeeProBot), unsupervised methods (DigitalDNA - B3 and B6), and non-contrastive self-supervised method (SATAR). 

\begin{table}[h]
  \centering
  \vspace{-0.20cm}
   \caption{Bot detection accuracy in terms of precision (\%), recall (\%), and F1-score, on two datasets with ground-truth labeled datasets. }
   \vspace{-0.30cm}
\label{tab:performance}
\scalebox{0.90}{
  \begin{tabular}{r:c c c c c c c}
   \toprule
   \multirow{2}{*}{\textbf{Method}}&\multirow{2}{*}{\textbf{Tier}}&\multicolumn{3}{c}{\textbf{Varol Dataset}}&\multicolumn{3}{c}{\textbf{Gilani Dataset}}\\ \cline{3-8}
  &&\textbf{Prec} & \textbf{Rec} &\textbf{F1}& \textbf{Prec} & \textbf{Rec} &\textbf{F1}\\
   \hline
   BotHunter & 0 & 67 & 61 & 61 & 73 & 70 & 70 \\
   DeeProBot & 0 & 52 & 51 & 50 & 47 & 48 & 47 \\
   DigitalDNA (B3) & 1 & 72 & 57 & 52 & 69 & 56 & 49 \\
   DigitalDNA (B6)& 1 & 75 & 56 & 50 & 69 & 57 & 52 \\
   Botometer ($\geq0.5$) & 2 & 75 & 75 & 75 & 73 & 73 & 73 \\
   Botometer ($\geq0.6$) & 2 & 75 & 75 & 75 & 71 & 71 & 71 \\
   Botometer ($\geq0.7$) & 2 & 75 & 72 & 73 & 70 & 69 & 69 \\
   SATAR & 2 & 58 & 50 & 40 & 64 & 60 & 58 \\ \hdashline
   BotSSCL & 0 & 70 & 73 & 70 & 74 & 74 & 74 \\
    BotSSCL(16$\times$64) & 1 & \textbf{80} & \textbf{81} & \textbf{80} & {78} & {78} & {78} \\
   BotSSCL(64$\times$64) & 1 & {76} & {78} & {77} & \textbf{79} & \textbf{79} & \textbf{79} \\
   \bottomrule
  \end{tabular}}
  \vspace{-0.4cm}
\end{table}

\subsubsection{Ablations:}
We now detail the importance of every design choice and hyperparameters in BotSSCL.

\textbf{Ablation Test 1--BotSSCL works better with data augmentation 1. } 
In this work, we tested three different data augmentation (abbreviated as Aug.) techniques to generate positives from anchor samples. We found that Aug. 1 achieves 80\% (F1) on Varol compared to 77\% and 78\% (F1) with Aug. 2 and 3, respectively. Similarly, Aug. 1 achieves 79\% (F1) on the Gilani dataset compared to 71\% and 77\% on Aug. 2 and 3, respectively. These results are present in  Appendix~\ref{appendix:performance} (Table~\ref{tab:all_evalutation_varol} and Table~\ref{tab:all_evalutation_gilani}). Our findings align with the fact that selecting the right augmentation choice is an essential strategy in contrastive learning as bias gets introduced through different Augmentation~\cite{jaiswal2020survey}. 

The reason behind the Augmentation 1 performance is that among the two encoders in BotSSCL, the optimal ones only extract relevant information about the contrastive task and throw away other irrelevant information to detect bots. In contrast, it is to be noted that \textit{MICE} imputations are computationally expensive due to the necessity of multiple iterations. This involves creating multiple copies of the data set, replacing the missing values with temporary placeholder values, and using regression models to predict missing NaN values. Thus, we set 30\% as the NaN replacement strategy to reduce the time needed to create augmentation. This means that 30\% of the data is imputed via MICE, and the remaining 70\% is the same between the anchor and the augmented version. Consequently, this highlights that we provided too much noise (more information between two views) rather than sufficient information required. Due to this, our model suffers from gradient bias rather than learning a high-level representation. In summary, our NaN replacement strategy limitation introduces the over-fitting of the shared mutual information, and hence, we achieve low performance with Augmentation 2.

On the other hand, linear transformation limits the direct overlapping between the anchor vector and the augmented view vector. This happens as it creates new data by linearly shifting the anchor vectors, creating a new representation (a positive sample). We envisage that Augmentation 3 performance is low because it has a risk that a negative sample's augmented data might become closer to the anchor instead of the augmented version (positive) due to the linear shift in the representation space. 

\begin{figure}[!ht]
  \centering
  \vspace{-0.35cm}
  \includegraphics[width=0.9\linewidth]{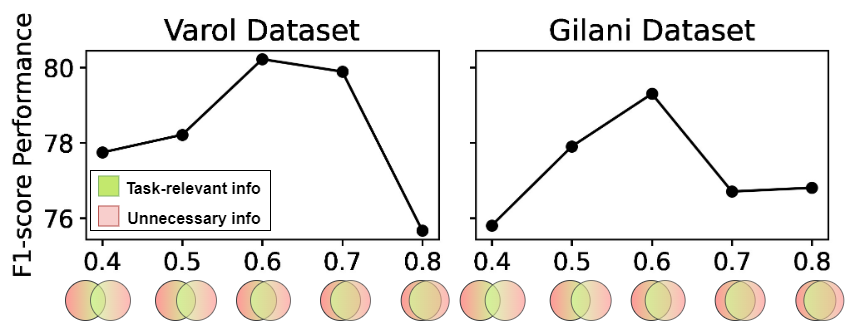}
  \vspace{-0.1cm}
  \caption{Mutual information variation with different corruption rates influences bot detection.
  }
  \label{fig:corruption_rate}
  \vspace{-0.3cm}
\end{figure}

\textbf{Ablation Test 2--The corruption rate in Augmentation 1 with 0.6 achieves the optimal representation for detection}. As mentioned, we achieved the best (F1) with augmentation 1. However, in augmentation 1, we can test different values of corruption rate. For instance, if the corruption rate is 0.4, it means 40\% of the data is corrupted in the anchor sample to generate a positive. Therefore, we tested with 0.4, 0.5, 0.6, 0.7, and 0.8 rates. As mentioned~(\S\ref{subsection:workingbotsscl}), a reverse U-shape will be achieved with an optimal corruption rate. We even achieve a reverse-U shape when we change corruption rate values, as shown in Figure~\ref{fig:corruption_rate}. Our findings suggest that 0.6 is the optimal corruption rate for learning representation for bot and human profiles (for precise values, please refer to Table~\ref{tab:corruption_rates}). 

\vspace{-0.3cm}
\begin{table}[!htb]
   \caption{Impact of corruption rates on F1-score (in \%) to reach the sweet spot for mutual information between views for the downstream task (i.e., classification of bots).} 
\label{tab:corruption_rates}
\vspace{-0.3cm}
  \begin{tabular}{c: c c}
   \toprule
  \textbf{Corruption Rate}& \textbf{Varol$_{(16\times64)}$}& \textbf{Gilani$_{(64\times64)}$}\\
   \hline
   0.4 & 77.7 & 75.8\\ 
   0.5 & 78.2 & 77.9\\
   \textbf{0.6} & \textbf{80.2} & \textbf{79.3}\\
   0.7 & 79.8 & 76.7\\
   0.8 & 75.6 & 76.8\\
   \bottomrule
  \end{tabular}
\end{table}

\vspace{-0.2cm}
\textbf{Ablation Test 3--BotSSCL works better with increasing the batch size}. 
For contrastive learning, we depend on the negatives that are drawn from the mini-batch. With a smaller batch size, only a few negatives are available. Thus, we tested the performance of BotSSCL with 128, 256, and 512 increasing values of batch sizes. Our findings suggest that a batch size of 512 achieves 76\% (F1) and 79\% (F1) on D1 and D2, respectively. In contrast, a batch size of 128 only achieves 75\% and 76\% (F1) on both datasets, respectively. The results are presented in Table~\ref{tab:batchsize1}. Our findings suggest that increasing batch size results in significant improvements, which are also highlighted in contrastive learning work like SimCLR~\cite{simclr}.
\vspace{-0.25cm}
\begin{table}[h]
   \caption{Impact of Batch Size in Model training in (\%)} 
   \label{tab:batchsize1}
   \vspace{-0.3cm}
\scalebox{0.90}{
 \begin{tabular}{c: c c c c}
   \toprule
  \textbf{Dataset}& \textbf{Batch Size}& \textbf{Prec} & \textbf{Rec} &\textbf{F1}\\
   \hline
   \multirow{3}{*}{Varol} & 128 & 74.6 & 75.8& 75.1 \\
   & 256 &73.8 & 75.9&74.5\\
   & \textbf{512} & \textbf{76.1}&\textbf{78.1} &\textbf{76.5}\\ \hdashline
   \multirow{3}{*}{Gilani} & 128 &76.1 &78.1 & 76.8 \\
   & 256 & 78.9& 78.8&76.8\\
   & \textbf{512} & \textbf{79.2}& \textbf{79.2}&\textbf{79.3}\\
   \bottomrule
  \end{tabular}}
\end{table}

\vspace{-0.15cm}
\textbf{Ablation Test 4--BotSSCL works better by increasing the number of epochs for training.} Similar to batch size, we can also increase the number of epochs for training the model. Generally, with more extended training, we expect BotSSCL to familiarize itself with various negatives drawn in different batches in different epochs. Therefore, we tested BotSSCL performance with 1000, 5000, and 10000 epochs, respectively. We see that performance is 76\% and 79\% on both datasets when the epoch is 5000 (see Table~\ref{tab:epochs}). However, performance drops when the epoch values are changed. 
\vspace{-0.15cm}
\begin{table}[!h]
   \caption{Impact of Epochs in (\%). Test configurations is 64 \textit{D} input dimension and 64 output embedding dimension.} 
\label{tab:epochs}
\vspace{-0.3cm}
\scalebox{0.88}{
  \begin{tabular}{c: c l c c c c}
   \toprule
  \textbf{Dataset}& \textbf{Epoch}& \textbf{Time Taken}& \textbf{Batch Size}& \textbf{Prec} & \textbf{Rec} &\textbf{F1}\\
   \hline
   \multirow{3}{*}{Varol} & 1,000 & $\approx 3$ minutes& 512 & 75.1& 77.2& 75.8 \\
   & \textbf{5,000}& $\approx 15$ minutes& \textbf{512} & \textbf{76.1}& \textbf{78.1}&\textbf{76.8}\\
   & 10,000& $\approx 30$ minutes& 512 & 75.1& 77.4&75.9\\ \hdashline
   \multirow{3}{*}{Gilani} & 1,000 &$\approx 3$ minutes& 512 & 77.1 & 76.5 & 76.7 \\
   & \textbf{5,000} &$\approx 15$ minutes & \textbf{512} &  \textbf{79.4} & \textbf{79.3}  & \textbf{79.4} \\
   & 10,000 & $\approx 30$ minutes& 512 & 78.3 & 78.1 & 78.2 \\
   \bottomrule
  \end{tabular}}
\end{table}

\textbf{Ablation Test 5--Alternatives to linear classifiers do not work better.}
We also evaluated the BotSSCL performance using different classifiers, such as linear SVC and a random forest (RF) model apart from logistic regression (LR). The reason to test representations using the random forest classifier is its widespread usage in current bot detection solutions~\cite{latah2020detection}. LR achieves $\approx 9\%$ and $\approx 5\%$ higher (F1) on both datasets, respectively, compared to RF. We see no reason to use any other classifier for the evaluation of representations from the results in ({Table~\ref{tab:differentclassifier}). We conjecture this is likely due to the small dataset size of Varol and Gilani, which resulted in a sparse tree in the random forest model. Theoretically, LR is faster and most commonly used in contrastive learning to evaluate and linearly separate representations in feature space. It is also a simple classification model, so we rely on it to evaluate BotSSCL's representation performance and generalization capability.
\begin{table}[!htb]
  \centering
  \vspace{-0.3cm}
   \caption{Bot detection performance in terms of prec. (\%), recall (\%), and F1-score, on two datasets with different classifiers.} 
\label{tab:differentclassifier}
\vspace{-0.3cm}
\scalebox{0.95}{
  \begin{tabular}{r: c c c: c c c: c c c}
   \toprule
   \multirow{2}{*}{\textbf{Dataset}}&\multicolumn{3}{c}{\textbf{Precision}}&\multicolumn{3}{c}{\textbf{Recall}}&\multicolumn{3}{c}{\textbf{F1-score}}\\ 
  &\textbf{LR} & \textbf{SVC} &\textbf{RF}&\textbf{LR} & \textbf{SVC} &\textbf{RF}&\textbf{LR} & \textbf{SVC} &\textbf{RF}\\
   \hline
   Varol & \textbf{76} & 78 & 75 & \textbf{77} & 74 & 69 & \textbf{77} & 75 & 70\\
   Gilani & \textbf{79}  & 79 & 77 & \textbf{79} & 78 & 75 & \textbf{79} & 78 & 75 \\
   \bottomrule
  \end{tabular}}
  \vspace{-0.4em}
\end{table}

\textbf{Ablation Test 6--Alternatives to InfoNCE loss do not work better.} We tested the importance of our choice of InfoNCE loss $(\mathcal{L}^{self})$ by replacing it with supervised loss $(\mathcal{L}^{sup})$ that is recently proposed in~\cite{khosla2020supervised}. We found self-supervised loss (in Eq. 7) to be $\approx 4\%$ and $\approx 5\%$ higher in terms of (F1) on both datasets, compared to supervised loss present in Equation~\ref{eq:10} as shown in Table~\ref{tab:loss}. The performance of $(\mathcal{L}^{sup})$ is low because the encoder network overfits the learning due to label information provided during training. 
\vspace{-0.2cm}
\begin{table}[!htb]
   \caption{Impact of variants of Contrastive Loss Functions in Model training (in \%)} 
\label{tab:loss}
\vspace{-0.3cm}
\scalebox{0.9}{
  \begin{tabular}{l: l c c c}
   \toprule
  \textbf{Dataset}& \textbf{Loss Function}& \textbf{Prec} & \textbf{Rec} &\textbf{F1}\\
   \hline
   \textbf{Varol$_{(16\times64)}$} & \textbf{Self-Supervised Loss ($\mathcal{L}^{self}$)} & \textbf{80}& \textbf{81}& \textbf{80} \\
   Varol$_{(16\times64)}$ & Supervised Loss ($\mathcal{L}^{sup}$) & 76&77 & 76 \\
   Varol$_{(16\times64)}$ & Supervised Loss ($\mathcal{L}^{sup}_{mod}$) & 77&77 & 77 \\ \hdashline
   \textbf{Gilani$_{(64\times64)}$} & \textbf{Self-Supervised Loss ($\mathcal{L}^{self}$)} & \textbf{79}& \textbf{79}& \textbf{79} \\
   Gilani$_{(64\times64)}$ & Supervised Loss ($\mathcal{L}^{sup}$) & 75& 74& 74 \\
   Gilani$_{(64\times64)}$ & Supervised Loss ($\mathcal{L}^{sup}_{mod}$) & 75& 75& 75 \\
   \bottomrule
  \end{tabular}}
\end{table}

\vspace{-0.4cm}
\subsection{Generalizability Test of BotSSCL (RQ2)}
\label{subsec:generalizability}
Generalizability aims to detect bots that are not present in the training data during the testing phase. This is an essential feature of BotSSCL as new bots can be activated at any time, and existing bots can exhibit adversarial behavior by making modifications to evade detection~\cite{Yang_Varol_Hui_Menczer_2020}. Also, different types of bots require different specializations for detection. For instance, a general-purpose bot detector faces a performance drop on cross-domain accounts (such as detecting retweeting patterns or anomalous behavior of bots)~\cite{cresci2023demystifying}. This is often seen as inappropriate use rather than a limitation of a bot detector. However, implementing multiple bot detectors does not help in terms of scalability for any OSN provider. 

To test how well our BotSSCL model generalizes on unseen data, we adopted the `Leave-one-botnet-out (LOBO)' methodology of Echeverr\'{i}a et al. \cite{lobo}. As the bots in our two datasets have different characteristics (shown in~\S\ref{subsec:performance} with varying model settings), we aim to assess whether BotSSCL can detect bots from different datasets (target class) by training on another dataset (training class). Table~\ref{tab:generalizability} shows all the generalizability experiments in terms of 1-class, 
and LOBO model F1-score. 1-class is when the model is trained and tested on the same dataset. 
The LOBO test is when the model is trained on other dataset and tested on the unseen target dataset. 

To make a comparison, we chose BotHunter (the second-best baseline) instead of Botometer (the first-best baseline) from Table~\ref{tab:performance}. This decision was influenced by the fact that Botometer is an API service already trained on both datasets, making it difficult to assess its generalizability. We observe in Table~\ref{tab:generalizability} that BotSSCL is generalizable for the D1 and D2 datasets. On D1 and D2, the LOBO model F1-score attained approximately 67\%, suggesting that BotSSCL can detect bots from both datasets with training on any of one dataset. This reduces the need for requiring training on the target-specific bot class. Hence, BotSSCL performs well on generalization. 

\vspace{-0.3cm}
\begin{table}[h]
   \caption{BotSSCL Generalizability performance in terms of 1-class, full model and LOBO model F1-score~(\%).} 
   \vspace{-0.20cm}
    \label{tab:generalizability}
    \scalebox{0.90}{
  \begin{tabular}{r:r c c c}
   \toprule
  \textbf{Method} & \textbf{Dataset}& \textbf{1-class F1}& 
  \textbf{LOBO model F1}\\
   \hline
   \multirow{2}{*}{\textbf{BotSSCL}}& Varol (D1) & 77 &  \textbf{68}\\ 
   & Gilani (D2) & 79 & \textbf{67} \\ \hdashline
   \multirow{2}{*}{\textbf{Bothunter}}& Varol (D1) & 61 & 51\\ 
   & Gilani (D2) & 70 & 47 \\
   \bottomrule
  \end{tabular}}
  \vspace{-0.5cm}
\end{table}

\subsection{Adversarial Robustness of BotSSCL (RQ3)}
\label{subsec:adversarial}
It is expected that once a bot detector is developed, its working details are shared widely. This openness allows the attacker to conduct evasion attacks~\cite{akhtar2023false}. In cybersecurity, evasion attacks involve successful attempts to deceive the model using unique deceptive inputs provided by the adversary~\cite{adversarialwebpage}.
In vision tasks, noise is added to the image (for example, a panda image) that is hard to identify through the human eye yet confuses the model's prediction (outputs as gibbon instead of panda)~\cite{goodfellow2017attacking}. In other words, verifying the successful adversarial image from human judgment is easy. On the other hand, in the numerical and tabular feature domain, subjective human judgment is a challenge. 

To succeed in the bot detection domain, an adversary can create a fake account with some modifications. Thus, adversaries must be considered when designing, proposing, or evaluating a bot detector. In this sub-section, we test our BotSSCL robustness to adversarial inputs, i.e., some of the combinations of modifications adversaries can carry in the Twitter profile. To test, we first define the scope of the adversary model's \textit{assumptions}, \textit{goals}, and \textit{capabilities}. This provides the environment for an adversary threat model for modeling possible attacks on BotSSCL. 

\textbf{Adversary Threat Model:} We conjecture that the adversary can act as a master bot handling multiple accounts using command and control channels. Specifically, we assume the adversary has a budget limit ($\approx\$1000$) of setting up or purchasing 1000 phone-verified fake accounts from various sources on the internet (known as NPC marketplace~\cite{buytwitteraccount}) or stolen accounts from underground forums or IRC channels~\cite{SHULMAN20105, stolenaccount, darkweb}. These budget and purchase limits are reasonable considering modest-size adversaries. In addition, we assume the adversary has black-box access to our BotSSCL model to use a trial-and-error process to evade the model. However, it should be noted that our BotSSCL depends on Twitter user representation~(\S\ref{subsec:twitteruserrep}) instead of using straightforward raw features extracted from profiles. Thus, the adversary has some knowledge of the four sub-modules used in the Twitter user representations and can only modify the features used in our framework.

An adversary, in our case, has two goals. First, to generate successful adversarial inputs to manipulate the model. Second, to limit the time it takes to generate a successful adversarial model to make it feasible to evasion attack. In other words, for our BotSSCL model to be robust to adversarial attacks, no adversary must exist that can create many accounts with non-negligible advantages. 

Finally, we define the adversary's capabilities in terms of modifications an adversary can do. Logically, an attacker can perform several modifications considering the four different sub-modules present in the Twitter user representation. The adversary can manipulate user metadata features, tweet metadata features, tweet temporal features, or all features together. 
The feature attributes used in perturbation for adversarial samples are based on seven prominent features. Two features belong to the \textit{user metadata feature}, i.e., the number of followers and the number of following count. Three features are from the \textit{Tweet metadata feature}, i.e. mean number of words in Tweet, mean number of favorites per Tweet, and mean number of retweets attained per Tweet. Lastly, two features, i.e., maximum number of Tweets per hour and maximum number of Tweets per day, are from the \textit{Tweet temporal feature}. For details, please refer to Appendix (\S\ref{appendix:adversarial}). 

We used a model named \textit{URET}~\cite{eykholt2023uret}  to generate an adversarial sample. URET requires a trained model (BotSSCL), dataset instances, and configuration files of the features with their possible perturbation values. Using a brute-force strategy, it visits all possible combinations of perturbations to dataset instances and tries to achieve a deceptive input for which the model flips its prediction. We only provided 200 samples (100 bot and human each) to URET for the adversarial generation process (please refer to Appendix~\ref{appendix:adversarial_choice} to know details for 200 samples choice). We provided the Gilani samples to URET as it already has bots closely resembling humans. 
In Table~\ref{tab:adversarial}, we observe that the \textit{success rate} (ratio of successful number of adversarial samples and 200 initial samples) for the adversarial sample is low for user metadata and tweet metadata feature manipulation, 
considering the time the adversary will have to spend. Similarly, an adversary only achieves a 4\% success rate if it tries to alter all features together. However, the sample generated by manipulating only temporal features achieves a 12.5\% success rate. While seemingly positive, this scenario does not favor bot adversaries since bots automatically and repeatedly have to post to promote their agenda. By altering the temporal aspects of Tweeting behavior, the adversary will find it challenging to achieve an artificial trend.

\begin{table}[!h]
  \centering
  \vspace{-0.25cm}
   \caption{Evaluation of Adversarial Robustness of BotSSCL in terms of \textit{success rate},
   number of \textit{adversarial samples} generated,
   and \textit{time taken} to brute force complete search space.}
   \vspace{-0.45cm}
\label{tab:adversarial}
\scalebox{0.835}{
  \begin{tabular}{l c c l}
   \toprule
   \multirow{2}{*}{\textbf{Adversarial Manipulation}}&\multicolumn{3}{c}{\textbf{BotSSCL (out of 200 samples)}}\\
   \cline{2-4}
  &\textbf{Success Rate} & \textbf{Samples} &\textbf{Time Taken}\\
   \hline
   1) User Metadata Feature  & 0.5\% & 1 & $\approx 10$ Hours \\
   2) Tweet Metadata Feature & 2.5 \% & 5 & $\approx 9$ Hours \\
   3) Tweet Temporal Feature  & 12.5 \% & 25 & $\approx 3$ Hours  \\ \hdashline
   4) All Above Three Feature  & 4.0 \% & 8 & $\approx 19.5$ Hours  \\
   \bottomrule
  \end{tabular}
  }
  \vspace{-0.45cm}
\end{table}

Moreover, we acknowledge that new-age adversaries can alter the tweets content and have this already in their toolbox. Thus, we have tested the robustness of our BotSSCL against text modification attacks of words in tweets separately, as infinite word space exists. We tested modifying the tweets using generative AI, particularly ChatGPT, a widely popular LLM. Moreover, we cannot modify the BERT embeddings of tweets randomly, as there is no way to ensure the correctness of the perturbed embeddings (the result might be meaningless). We also incorporated changes in the tweet metadata features after tweet content perturbations, representing more accurate perturbations. Our analysis in Appendix~\ref{appendix:adversarial_gpt} shows that textual adversarial attack
success is at most 5\%.

\subsection{Discussion}
The proposed BotSSCL framework highlights three key aspects: \textit{(1)} a comprehensive performance evaluation on datasets comprising sophisticated bots imitating humans, \textit{(2)} providing generalizability by learning only task-relevant information, and \textit{(3)} resiliency against adversarial attacks.
We think that established services, such as Botometer~\cite{botometer}, which train new machine-learning models in response to the recent Twitter API changes, can take advantage of our framework.
Our results reveal some key findings, listed below: 
\begin{enumerate}[wide, labelwidth=!, labelindent=0pt]
    \item We found that bot detection configurations need to be tailored to specific datasets, likely due to variations in bot characteristics across different datasets. 
    Since bots in the Gilani dataset had lower homogeneity (KNN cluster homogeneity score) than bots in the Varol dataset, we found that the Gilani dataset requires a larger input dimension (i.e., more information) to distinguish. between bots and humans.
    \item  Despite the different configurations needed, even with a generalized configuration (64x64), BotSSCL outperforms SOTA on these two datasets. Our BotSSCL is generalizable as the quality of our learned representations shows that the encoder network only learned task-relevant information (i.e., detecting bots) and threw out irrelevant information (which is noise). 
    Further, we show six ablations on BotSSCL by altering hyperparameters. We observe that increasing batch size and epochs in contrastive learning helps. 
    \item Due to Twitter user representation, attackers will have more difficulty conducting adversarial attacks. To succeed, an attacker must modify all feature categories, especially Tweet's temporal activity. We found that if an attacker modifies Tweet's temporal activity in the profile by tweeting at a regular rate like humans, this will contradict the adversary's goal of creating an artificial trend.
    \item 
    Our Twitter user representation does not rely on direct raw features but instead on a combination of user representation vectors, as it is also a case that frequent changes occur in the platform's API's endpoints. Considering the rapid changes, it is normal that a few features do not remain accessible after API changes, such as followers growth rate per tweet~\cite{dimitriadis2021social}. Thus, utilizing user representation is beneficial.
\end{enumerate}

\textbf{Limitations:} To create a generalized model, we have used one standard setting 64x64 for both datasets. This reduces performance on the Varol dataset. We have not investigated which particular feature set is more critical in the decision-making of both datasets.  Moreover, our work used the basic MICE imputation method to fill the NaN instance in the dataset features. This can be replaced with recent generative approaches that can recover feature values by approximating feature distribution. The same generative models can help create synthetic real-world augmentation for Twitter accounts. 
Moreover, in the future, we will use newly collected datasets such as Twi-bot 20 and Twi-bot 22 which may lead to a better understanding of contrastive learning for bot detection. 

\vspace{-0.5em}
\section{Related Work}
\label{sec:related_work}
\textbf{Social Bot Detection:} Several works have proposed machine learning models to detect bot accounts on Twitter~\cite{botometer,bothunter,digitaldna,satar,deeprobot}.
The proposed models include random forests, decision trees, deep learning, and neural networks, all employing various feature inputs from five categories, as indicated in Table~\ref{tab:features_used}. 
DigitalDNA uses a learning classifier system to break types of tweets and their interactions into specialized string sequences, which was a creative approach for a light implementation~\cite{digitaldna}. SATAR, a self-supervised model, uses all features to build a model that weighs more heavily on the network interactions and an account relationship to other accounts~\cite{satar}. 
All the above methods face issues with classifying sophisticated bots that mimic humans. Moreover, the above methods have not shown any test for generalizability or adversarial robustness. Our proposed BotSSCL framework improves performance by using contrastive learning, a self-supervised paradigm to discriminate between samples. To our knowledge, no other work has employed self-supervised contrastive learning in social bot detection. We show the guarantees for generalizability and robustness against adversarial attacks. In addition, our model uses features under Tier 1, which is feasible to collect under API constraints.

\textbf{Contrastive Learning: }Deep learning algorithms have achieved state-of-the-art performance in many computer vision tasks~\cite{jing2020self,zhai2019s4l, simclr}. However, little has been explored on a real-world \textit{tabular} dataset, which does not have the same structure as image and language data. 
Initial work such as VIME~\cite{yoon2020vime} uses feature and mask estimation, which requires the knowledge of a pretext task. 
On the other hand, a few methods have recently used contrastive learning on tabular data but differ in how they generate views. For instance, 
SAINT~\cite{somepalli2021saint} uses CutMix~\cite{yun2019cutmix} in the input space and mixup~\cite{zhang2017mixup} in the embedding space to generate different views. Another approach 
LocL framework~\cite{gharibshah2022local} leverages feature correlation to assign strongly correlated features next to each other for local learning. 
The assumption is that instead of using all features, feature subsets by meaningful reordering can help CNN kernels capture spatial connections.
Similarly, TabNet~\cite{arik2021tabnet} uses sequential attention for selecting essential features, and TaBERT~\cite{yin2020tabert} encodes a subset of content most relevant to input utterance. 
In contrast, SubTab~\cite{ucar2021subtab} divides the features into multiple subsets similar to cropping images and uses an MLP-based autoencoder for representation learning better than the CNN-based model. On the other hand, CBD~\cite{zhou2023detecting} employs a two-stage learning—first, contrastive learning, and second, fine-tuning using graph-based techniques. The drawbacks of the previous methods are relying on feature correlation or reconstruction loss to recover the original data from corrupted data, which restricts the computational scalability. BotSSCL is computationally scalable due to reliance on only a single contrastive loss. Our GNN-free framework does not require dependency on relational data such as the `follow relationship', decreasing data collection capacity, complexity, and additional API calls. Additionally, our MLP-based encoders are similar to SCARF~\cite{bahri2021scarf}
but differs in the depth sizes of the encoder and projection head configuration and two additional novel augmentation techniques. The MLP-based encoder performs better than the CNN-based or GNN-based models, as this work does not assume feature correlation exists in tabular datasets to capture spatial relationships. 

\section{Conclusion}
\label{sec:conclusion}
This paper proposes BotSSCL, a framework to detect social bots that mimic genuine OSN users. We use a novel framework of contrastive learning that differentiates between bot and human clusters to improve linear separability. The BotSSCL outperformed SOTA models with $\approx6\%$ and $\approx8\%$ higher (F1) on two tabular feature bot datasets, respectively. Additionally, we highlighted the performance of BotSSCL under six ablation tests. Our findings suggest that our framework is generalizable and robust to feature manipulation.

\bibliographystyle{ACM-Reference-Format}
\bibliography{sample-base}


\begin{thebibliography}{69}


\ifx \showCODEN    \undefined \def \showCODEN     #1{\unskip}     \fi
\ifx \showDOI      \undefined \def \showDOI       #1{#1}\fi
\ifx \showISBNx    \undefined \def \showISBNx     #1{\unskip}     \fi
\ifx \showISBNxiii \undefined \def \showISBNxiii  #1{\unskip}     \fi
\ifx \showISSN     \undefined \def \showISSN      #1{\unskip}     \fi
\ifx \showLCCN     \undefined \def \showLCCN      #1{\unskip}     \fi
\ifx \shownote     \undefined \def \shownote      #1{#1}          \fi
\ifx \showarticletitle \undefined \def \showarticletitle #1{#1}   \fi
\ifx \showURL      \undefined \def \showURL       {\relax}        \fi
\providecommand\bibfield[2]{#2}
\providecommand\bibinfo[2]{#2}
\providecommand\natexlab[1]{#1}
\providecommand\showeprint[2][]{arXiv:#2}

\bibitem[Akhtar et~al\mbox{.}(2023a)]%
        {majidlcn}
\bibfield{author}{\bibinfo{person}{Mohammad~Majid Akhtar}, \bibinfo{person}{Ishan Karunanayake}, \bibinfo{person}{Bibhas Sharma}, \bibinfo{person}{Rahat Masood}, \bibinfo{person}{Muhammad Ikram}, {and} \bibinfo{person}{Salil~S. Kanhere}.} \bibinfo{year}{2023}\natexlab{a}.
\newblock \showarticletitle{Towards Automatic Annotation and Detection of Fake News}. In \bibinfo{booktitle}{\emph{2023 IEEE 48th Conference on Local Computer Networks (LCN)}}. \bibinfo{pages}{1--9}.
\newblock
\urldef\tempurl%
\url{https://doi.org/10.1109/LCN58197.2023.10223359}
\showDOI{\tempurl}


\bibitem[Akhtar et~al\mbox{.}(2023b)]%
        {akhtar2023false}
\bibfield{author}{\bibinfo{person}{Mohammad~Majid Akhtar}, \bibinfo{person}{Rahat Masood}, \bibinfo{person}{Muhammad Ikram}, {and} \bibinfo{person}{Salil~S. Kanhere}.} \bibinfo{year}{2023}\natexlab{b}.
\newblock \bibinfo{title}{False Information, Bots and Malicious Campaigns: Demystifying Elements of Social Media Manipulations}.
\newblock
\newblock
\showeprint[arxiv]{2308.12497}~[cs.SI]


\bibitem[Arik and Pfister(2021)]%
        {arik2021tabnet}
\bibfield{author}{\bibinfo{person}{Sercan~{\"O} Arik} {and} \bibinfo{person}{Tomas Pfister}.} \bibinfo{year}{2021}\natexlab{}.
\newblock \showarticletitle{Tabnet: Attentive interpretable tabular learning}. In \bibinfo{booktitle}{\emph{Proceedings of the AAAI conference on artificial intelligence}}, Vol.~\bibinfo{volume}{35}. \bibinfo{pages}{6679--6687}.
\newblock


\bibitem[Arya(2022)]%
        {adversarialwebpage}
\bibfield{author}{\bibinfo{person}{Nisha Arya}.} \bibinfo{year}{2022}\natexlab{}.
\newblock \bibinfo{title}{What is Adversarial Machine Learning?}
\newblock \bibinfo{howpublished}{\url{https://www.kdnuggets.com/2022/03/adversarial-machine-learning.html}}.
\newblock
\newblock
\shownote{Accessed: 19 September 2023}.


\bibitem[Bahri et~al\mbox{.}(2021)]%
        {bahri2021scarf}
\bibfield{author}{\bibinfo{person}{Dara Bahri}, \bibinfo{person}{Heinrich Jiang}, \bibinfo{person}{Yi Tay}, {and} \bibinfo{person}{Donald Metzler}.} \bibinfo{year}{2021}\natexlab{}.
\newblock \showarticletitle{Scarf: Self-supervised contrastive learning using random feature corruption}.
\newblock \bibinfo{journal}{\emph{arXiv preprint arXiv:2106.15147}} (\bibinfo{year}{2021}).
\newblock


\bibitem[Beskow and Carley(2018)]%
        {bothunter}
\bibfield{author}{\bibinfo{person}{David Beskow} {and} \bibinfo{person}{Kathleen Carley}.} \bibinfo{year}{2018}\natexlab{}.
\newblock \showarticletitle{Bot-hunter: A Tiered Approach to Detecting \& Characterizing Automated Activity on Twitter}.
\newblock


\bibitem[Beskow and Carley(2019)]%
        {beskow2019its}
\bibfield{author}{\bibinfo{person}{David~M Beskow} {and} \bibinfo{person}{Kathleen~M Carley}.} \bibinfo{year}{2019}\natexlab{}.
\newblock \showarticletitle{Its all in a name: detecting and labeling bots by their name}.
\newblock \bibinfo{journal}{\emph{Computational and mathematical organization theory}}  \bibinfo{volume}{25} (\bibinfo{year}{2019}), \bibinfo{pages}{24--35}.
\newblock


\bibitem[Botometer(2023)]%
        {botometerrepository}
\bibfield{author}{\bibinfo{person}{Botometer}.} \bibinfo{year}{2023}\natexlab{}.
\newblock \bibinfo{title}{Bot Repository}.
\newblock \bibinfo{howpublished}{\url{https://botometer.osome.iu.edu/bot-repository/datasets.html}}.
\newblock
\newblock
\shownote{Accessed: 1 March 2023}.


\bibitem[Chen et~al\mbox{.}(2020)]%
        {simclr}
\bibfield{author}{\bibinfo{person}{Ting Chen}, \bibinfo{person}{Simon Kornblith}, \bibinfo{person}{Mohammad Norouzi}, {and} \bibinfo{person}{Geoffrey Hinton}.} \bibinfo{year}{2020}\natexlab{}.
\newblock \showarticletitle{A Simple Framework for Contrastive Learning of Visual Representations}. In \bibinfo{booktitle}{\emph{Proceedings of the 37th International Conference on Machine Learning}} \emph{(\bibinfo{series}{ICML'20})}. \bibinfo{publisher}{JMLR.org}, Article \bibinfo{articleno}{149}, \bibinfo{numpages}{11}~pages.
\newblock


\bibitem[Cresci(2020a)]%
        {cresci2020decade}
\bibfield{author}{\bibinfo{person}{Stefano Cresci}.} \bibinfo{year}{2020}\natexlab{a}.
\newblock \showarticletitle{A decade of social bot detection}.
\newblock \bibinfo{journal}{\emph{Commun. ACM}} \bibinfo{volume}{63}, \bibinfo{number}{10} (\bibinfo{year}{2020}), \bibinfo{pages}{72--83}.
\newblock


\bibitem[Cresci(2020b)]%
        {cresci2020detecting}
\bibfield{author}{\bibinfo{person}{Stefano Cresci}.} \bibinfo{year}{2020}\natexlab{b}.
\newblock \showarticletitle{Detecting malicious social bots: story of a never-ending clash}. In \bibinfo{booktitle}{\emph{Disinformation in Open Online Media: First Multidisciplinary International Symposium, MISDOOM 2019, Hamburg, Germany, February 27--March 1, 2019, Revised Selected Papers 1}}. Springer, \bibinfo{pages}{77--88}.
\newblock


\bibitem[Cresci et~al\mbox{.}(2021)]%
        {cresci2021coming}
\bibfield{author}{\bibinfo{person}{Stefano Cresci}, \bibinfo{person}{Marinella Petrocchi}, \bibinfo{person}{Angelo Spognardi}, {and} \bibinfo{person}{Stefano Tognazzi}.} \bibinfo{year}{2021}\natexlab{}.
\newblock \showarticletitle{The coming age of adversarial social bot detection}.
\newblock \bibinfo{journal}{\emph{First Monday}} (\bibinfo{year}{2021}).
\newblock


\bibitem[Cresci et~al\mbox{.}(2018)]%
        {digitaldna}
\bibfield{author}{\bibinfo{person}{Stefano Cresci}, \bibinfo{person}{Roberto~Di Pietro}, \bibinfo{person}{Marinella Petrocchi}, \bibinfo{person}{Angelo Spognardi}, {and} \bibinfo{person}{Maurizio Tesconi}.} \bibinfo{year}{2018}\natexlab{}.
\newblock \showarticletitle{Social Fingerprinting: Detection of Spambot Groups Through DNA-Inspired Behavioral Modeling}.
\newblock \bibinfo{journal}{\emph{IEEE Transactions on Dependable and Secure Computing}} \bibinfo{volume}{15}, \bibinfo{number}{4} (\bibinfo{year}{2018}), \bibinfo{pages}{561--576}.
\newblock
\urldef\tempurl%
\url{https://doi.org/10.1109/TDSC.2017.2681672}
\showDOI{\tempurl}


\bibitem[Cresci et~al\mbox{.}(2023)]%
        {cresci2023demystifying}
\bibfield{author}{\bibinfo{person}{Stefano Cresci}, \bibinfo{person}{Roberto~Di Pietro}, \bibinfo{person}{Angelo Spognardi}, \bibinfo{person}{Maurizio Tesconi}, {and} \bibinfo{person}{Marinella Petrocchi}.} \bibinfo{year}{2023}\natexlab{}.
\newblock \bibinfo{title}{Demystifying Misconceptions in Social Bots Research}.
\newblock
\newblock
\showeprint[arxiv]{2303.17251}~[cs.SI]


\bibitem[Dimitriadis et~al\mbox{.}(2021)]%
        {dimitriadis2021social}
\bibfield{author}{\bibinfo{person}{Ilias Dimitriadis}, \bibinfo{person}{Konstantinos Georgiou}, {and} \bibinfo{person}{Athena Vakali}.} \bibinfo{year}{2021}\natexlab{}.
\newblock \showarticletitle{Social botomics: A systematic ensemble ml approach for explainable and multi-class bot detection}.
\newblock \bibinfo{journal}{\emph{Applied Sciences}} \bibinfo{volume}{11}, \bibinfo{number}{21} (\bibinfo{year}{2021}), \bibinfo{pages}{9857}.
\newblock


\bibitem[Echeverr\"{\i}£¡a et~al\mbox{.}(2018)]%
        {lobo}
\bibfield{author}{\bibinfo{person}{Juan Echeverr\"{\i}£¡a}, \bibinfo{person}{Emiliano De~Cristofaro}, \bibinfo{person}{Nicolas Kourtellis}, \bibinfo{person}{Ilias Leontiadis}, \bibinfo{person}{Gianluca Stringhini}, {and} \bibinfo{person}{Shi Zhou}.} \bibinfo{year}{2018}\natexlab{}.
\newblock \showarticletitle{LOBO: Evaluation of Generalization Deficiencies in Twitter Bot Classifiers}. In \bibinfo{booktitle}{\emph{Proceedings of the 34th Annual Computer Security Applications Conference}} (San Juan, PR, USA) \emph{(\bibinfo{series}{ACSAC '18})}. \bibinfo{publisher}{Association for Computing Machinery}, \bibinfo{address}{New York, NY, USA}, \bibinfo{pages}{137–146}.
\newblock
\showISBNx{9781450365697}
\urldef\tempurl%
\url{https://doi.org/10.1145/3274694.3274738}
\showDOI{\tempurl}


\bibitem[Eykholt et~al\mbox{.}(2023)]%
        {eykholt2023uret}
\bibfield{author}{\bibinfo{person}{Kevin Eykholt}, \bibinfo{person}{Taesung Lee}, \bibinfo{person}{Douglas Schales}, \bibinfo{person}{Jiyong Jang}, {and} \bibinfo{person}{Ian Molloy}.} \bibinfo{year}{2023}\natexlab{}.
\newblock \showarticletitle{$\{$URET$\}$: Universal Robustness Evaluation Toolkit (for Evasion)}. In \bibinfo{booktitle}{\emph{32nd USENIX Security Symposium (USENIX Security 23)}}. \bibinfo{pages}{3817--3833}.
\newblock


\bibitem[Feng et~al\mbox{.}(2021)]%
        {satar}
\bibfield{author}{\bibinfo{person}{Shangbin Feng}, \bibinfo{person}{Herun Wan}, \bibinfo{person}{Ningnan Wang}, \bibinfo{person}{Jundong Li}, {and} \bibinfo{person}{Minnan Luo}.} \bibinfo{year}{2021}\natexlab{}.
\newblock \showarticletitle{{SATAR}}. In \bibinfo{booktitle}{\emph{Proceedings of the 30th {ACM} International Conference on Information \& Knowledge Management}}. \bibinfo{publisher}{{ACM}}.
\newblock
\urldef\tempurl%
\url{https://doi.org/10.1145/3459637.3481949}
\showDOI{\tempurl}


\bibitem[Ferrara(2023)]%
        {ferrara2023social}
\bibfield{author}{\bibinfo{person}{Emilio Ferrara}.} \bibinfo{year}{2023}\natexlab{}.
\newblock \showarticletitle{Social bot detection in the age of ChatGPT: Challenges and opportunities}.
\newblock \bibinfo{journal}{\emph{First Monday}} (\bibinfo{year}{2023}).
\newblock


\bibitem[(formerly Twitter)(2023a)]%
        {twitterratelimit}
\bibfield{author}{\bibinfo{person}{X (formerly Twitter)}.} \bibinfo{year}{2023}\natexlab{a}.
\newblock \bibinfo{title}{About Twitter limits}.
\newblock \bibinfo{howpublished}{\url{https://tinyurl.com/tweet-rate-limits/}}.
\newblock
\newblock
\shownote{Accessed: 18 September 2023}.


\bibitem[(formerly Twitter)(2023b)]%
        {tweetperhour}
\bibfield{author}{\bibinfo{person}{X (formerly Twitter)}.} \bibinfo{year}{2023}\natexlab{b}.
\newblock \bibinfo{title}{Rate limits: Standard v1.1}.
\newblock \bibinfo{howpublished}{\url{https://tinyurl.com/tweet-per-hour/}}.
\newblock
\newblock
\shownote{Accessed: 18 September 2023}.


\bibitem[Gharibshah and Zhu(2022)]%
        {gharibshah2022local}
\bibfield{author}{\bibinfo{person}{Zhabiz Gharibshah} {and} \bibinfo{person}{Xingquan Zhu}.} \bibinfo{year}{2022}\natexlab{}.
\newblock \showarticletitle{Local Contrastive Feature Learning for Tabular Data}. In \bibinfo{booktitle}{\emph{Proceedings of the 31st ACM International Conference on Information \& Knowledge Management}}. \bibinfo{pages}{3963--3967}.
\newblock


\bibitem[Gilani et~al\mbox{.}(2017)]%
        {gilani}
\bibfield{author}{\bibinfo{person}{Zafar Gilani}, \bibinfo{person}{Reza Farahbakhsh}, \bibinfo{person}{Gareth Tyson}, \bibinfo{person}{Liang Wang}, {and} \bibinfo{person}{Jon Crowcroft}.} \bibinfo{year}{2017}\natexlab{}.
\newblock \showarticletitle{Of Bots and Humans (on Twitter)}. In \bibinfo{booktitle}{\emph{Proceedings of the 2017 IEEE/ACM International Conference on Advances in Social Networks Analysis and Mining 2017}} (Sydney, Australia) \emph{(\bibinfo{series}{ASONAM '17})}. \bibinfo{publisher}{Association for Computing Machinery}, \bibinfo{address}{New York, NY, USA}, \bibinfo{pages}{349–354}.
\newblock
\showISBNx{9781450349932}
\urldef\tempurl%
\url{https://doi.org/10.1145/3110025.3110090}
\showDOI{\tempurl}


\bibitem[Goodfellow et~al\mbox{.}(2017)]%
        {goodfellow2017attacking}
\bibfield{author}{\bibinfo{person}{Ian Goodfellow}, \bibinfo{person}{Nicolas Papernot}, \bibinfo{person}{Sandy Huang}, \bibinfo{person}{Yan Duan}, \bibinfo{person}{Pieter Abbeel}, {and} \bibinfo{person}{Jack Clark}.} \bibinfo{year}{2017}\natexlab{}.
\newblock \showarticletitle{Attacking machine learning with adversarial examples}.
\newblock \bibinfo{journal}{\emph{OpenAI Blog}}  \bibinfo{volume}{24} (\bibinfo{year}{2017}).
\newblock


\bibitem[Guo et~al\mbox{.}(2021)]%
        {guo2021social}
\bibfield{author}{\bibinfo{person}{Qinglang Guo}, \bibinfo{person}{Haiyong Xie}, \bibinfo{person}{Yangyang Li}, \bibinfo{person}{Wen Ma}, {and} \bibinfo{person}{Chao Zhang}.} \bibinfo{year}{2021}\natexlab{}.
\newblock \showarticletitle{Social bots detection via fusing bert and graph convolutional networks}.
\newblock \bibinfo{journal}{\emph{Symmetry}} \bibinfo{volume}{14}, \bibinfo{number}{1} (\bibinfo{year}{2021}), \bibinfo{pages}{30}.
\newblock


\bibitem[Haixiang et~al\mbox{.}(2017)]%
        {haixiang2017learning}
\bibfield{author}{\bibinfo{person}{Guo Haixiang}, \bibinfo{person}{Li Yijing}, \bibinfo{person}{Jennifer Shang}, \bibinfo{person}{Gu Mingyun}, \bibinfo{person}{Huang Yuanyue}, {and} \bibinfo{person}{Gong Bing}.} \bibinfo{year}{2017}\natexlab{}.
\newblock \showarticletitle{Learning from class-imbalanced data: Review of methods and applications}.
\newblock \bibinfo{journal}{\emph{Expert systems with applications}}  \bibinfo{volume}{73} (\bibinfo{year}{2017}), \bibinfo{pages}{220--239}.
\newblock


\bibitem[Hayawi et~al\mbox{.}(2022)]%
        {deeprobot}
\bibfield{author}{\bibinfo{person}{Kadhim Hayawi}, \bibinfo{person}{Sujith Mathew}, \bibinfo{person}{Neethu Venugopal}, \bibinfo{person}{Mohammad~M. Masud}, {and} \bibinfo{person}{Pin-Han Ho}.} \bibinfo{year}{2022}\natexlab{}.
\newblock \showarticletitle{DeeProBot: a hybrid deep neural network model for social bot detection based on user profile data}.
\newblock \bibinfo{journal}{\emph{Social Network Analysis and Mining}} \bibinfo{volume}{12}, \bibinfo{number}{1} (\bibinfo{date}{12 Mar} \bibinfo{year}{2022}), \bibinfo{pages}{43}.
\newblock
\showISSN{1869-5469}
\urldef\tempurl%
\url{https://doi.org/10.1007/s13278-022-00869-w}
\showDOI{\tempurl}


\bibitem[India(2023)]%
        {buytwitteraccount}
\bibfield{author}{\bibinfo{person}{Outlook India}.} \bibinfo{year}{2023}\natexlab{}.
\newblock \bibinfo{title}{Buy Twitter Accounts From The Top Sites In 2023}.
\newblock \bibinfo{howpublished}{\url{https://www.outlookindia.com/business-spotlight/buy-twitter-accounts-from-the-top-sites-in-2023-news-267701}}.
\newblock
\newblock
\shownote{Accessed: 19 September 2023}.


\bibitem[Jaiswal et~al\mbox{.}(2020)]%
        {jaiswal2020survey}
\bibfield{author}{\bibinfo{person}{Ashish Jaiswal}, \bibinfo{person}{Ashwin~Ramesh Babu}, \bibinfo{person}{Mohammad~Zaki Zadeh}, \bibinfo{person}{Debapriya Banerjee}, {and} \bibinfo{person}{Fillia Makedon}.} \bibinfo{year}{2020}\natexlab{}.
\newblock \showarticletitle{A survey on contrastive self-supervised learning}.
\newblock \bibinfo{journal}{\emph{Technologies}} \bibinfo{volume}{9}, \bibinfo{number}{1} (\bibinfo{year}{2020}), \bibinfo{pages}{2}.
\newblock


\bibitem[Jing and Tian(2020)]%
        {jing2020self}
\bibfield{author}{\bibinfo{person}{Longlong Jing} {and} \bibinfo{person}{Yingli Tian}.} \bibinfo{year}{2020}\natexlab{}.
\newblock \showarticletitle{Self-supervised visual feature learning with deep neural networks: A survey}.
\newblock \bibinfo{journal}{\emph{IEEE transactions on pattern analysis and machine intelligence}} \bibinfo{volume}{43}, \bibinfo{number}{11} (\bibinfo{year}{2020}), \bibinfo{pages}{4037--4058}.
\newblock


\bibitem[Khosla et~al\mbox{.}(2020)]%
        {khosla2020supervised}
\bibfield{author}{\bibinfo{person}{Prannay Khosla}, \bibinfo{person}{Piotr Teterwak}, \bibinfo{person}{Chen Wang}, \bibinfo{person}{Aaron Sarna}, \bibinfo{person}{Yonglong Tian}, \bibinfo{person}{Phillip Isola}, \bibinfo{person}{Aaron Maschinot}, \bibinfo{person}{Ce Liu}, {and} \bibinfo{person}{Dilip Krishnan}.} \bibinfo{year}{2020}\natexlab{}.
\newblock \showarticletitle{Supervised contrastive learning}.
\newblock \bibinfo{journal}{\emph{Advances in neural information processing systems}}  \bibinfo{volume}{33} (\bibinfo{year}{2020}), \bibinfo{pages}{18661--18673}.
\newblock


\bibitem[Latah(2020)]%
        {latah2020detection}
\bibfield{author}{\bibinfo{person}{Majd Latah}.} \bibinfo{year}{2020}\natexlab{}.
\newblock \showarticletitle{Detection of malicious social bots: A survey and a refined taxonomy}.
\newblock \bibinfo{journal}{\emph{Expert Systems with Applications}}  \bibinfo{volume}{151} (\bibinfo{year}{2020}), \bibinfo{pages}{113383}.
\newblock


\bibitem[Lombardi et~al\mbox{.}(2022)]%
        {lombardi2022ai}
\bibfield{author}{\bibinfo{person}{Flavio Lombardi}, \bibinfo{person}{Maurantonio Caprolu}, {and} \bibinfo{person}{Roberto Di~Pietro}.} \bibinfo{year}{2022}\natexlab{}.
\newblock \showarticletitle{AI-enabled Bot and Social Media: A Survey of Tools, Techniques, and Platforms for the Arms Race}.
\newblock \bibinfo{journal}{\emph{Mixed Methods Perspectives on Communication and Social Media Research}} (\bibinfo{year}{2022}), \bibinfo{pages}{255--269}.
\newblock


\bibitem[Lones(2021)]%
        {lones2021avoid}
\bibfield{author}{\bibinfo{person}{Michael~A Lones}.} \bibinfo{year}{2021}\natexlab{}.
\newblock \showarticletitle{How to avoid machine learning pitfalls: a guide for academic researchers}.
\newblock \bibinfo{journal}{\emph{arXiv preprint arXiv:2108.02497}} (\bibinfo{year}{2021}).
\newblock


\bibitem[Madahali and Hall(2020)]%
        {benfordlaw}
\bibfield{author}{\bibinfo{person}{Lale Madahali} {and} \bibinfo{person}{Margeret Hall}.} \bibinfo{year}{2020}\natexlab{}.
\newblock \showarticletitle{Application of the Benford’s law to Social bots and Information Operations activities}. In \bibinfo{booktitle}{\emph{2020 International Conference on Cyber Situational Awareness, Data Analytics and Assessment (CyberSA)}}. \bibinfo{pages}{1--8}.
\newblock
\urldef\tempurl%
\url{https://doi.org/10.1109/CyberSA49311.2020.9139709}
\showDOI{\tempurl}


\bibitem[McMillan(2010)]%
        {stolenaccount}
\bibfield{author}{\bibinfo{person}{Robert McMillan}.} \bibinfo{year}{2010}\natexlab{}.
\newblock \bibinfo{title}{Stolen Twitter accounts can fetch \$1,000}.
\newblock \bibinfo{howpublished}{\url{https://www.computerworld.com/article/2760814/stolen-twitter-accounts-can-fetch--1-000.html}}.
\newblock
\newblock
\shownote{Accessed: 19 September 2023}.


\bibitem[Morris et~al\mbox{.}(2020)]%
        {morris2020textattack}
\bibfield{author}{\bibinfo{person}{John~X Morris}, \bibinfo{person}{Eli Lifland}, \bibinfo{person}{Jin~Yong Yoo}, \bibinfo{person}{Jake Grigsby}, \bibinfo{person}{Di Jin}, {and} \bibinfo{person}{Yanjun Qi}.} \bibinfo{year}{2020}\natexlab{}.
\newblock \showarticletitle{Textattack: A framework for adversarial attacks, data augmentation, and adversarial training in nlp}.
\newblock \bibinfo{journal}{\emph{arXiv preprint arXiv:2005.05909}} (\bibinfo{year}{2020}).
\newblock


\bibitem[Mou and Lee(2020)]%
        {mou2020malicious}
\bibfield{author}{\bibinfo{person}{Guanyi Mou} {and} \bibinfo{person}{Kyumin Lee}.} \bibinfo{year}{2020}\natexlab{}.
\newblock \showarticletitle{Malicious bot detection in online social networks: arming handcrafted features with deep learning}. In \bibinfo{booktitle}{\emph{Social Informatics: 12th International Conference, SocInfo 2020, Pisa, Italy, October 6--9, 2020, Proceedings 12}}. Springer, \bibinfo{pages}{220--236}.
\newblock


\bibitem[OpenAI(2023)]%
        {chatgpt}
\bibfield{author}{\bibinfo{person}{OpenAI}.} \bibinfo{year}{2023}\natexlab{}.
\newblock \bibinfo{title}{ChatGPT}.
\newblock \bibinfo{howpublished}{\url{https://chat.openai.com/}}.
\newblock
\newblock
\shownote{Accessed: 1 November 2023}.


\bibitem[Pedregosa et~al\mbox{.}(2011)]%
        {pedregosa2011scikit}
\bibfield{author}{\bibinfo{person}{Fabian Pedregosa}, \bibinfo{person}{Ga{\"e}l Varoquaux}, \bibinfo{person}{Alexandre Gramfort}, \bibinfo{person}{Vincent Michel}, \bibinfo{person}{Bertrand Thirion}, \bibinfo{person}{Olivier Grisel}, \bibinfo{person}{Mathieu Blondel}, \bibinfo{person}{Peter Prettenhofer}, \bibinfo{person}{Ron Weiss}, \bibinfo{person}{Vincent Dubourg}, {et~al\mbox{.}}} \bibinfo{year}{2011}\natexlab{}.
\newblock \showarticletitle{Scikit-learn: Machine learning in Python}.
\newblock \bibinfo{journal}{\emph{Journal of machine learning research}} \bibinfo{volume}{12}, \bibinfo{number}{Oct} (\bibinfo{year}{2011}), \bibinfo{pages}{2825--2830}.
\newblock


\bibitem[QuillBot(2023)]%
        {quillbot}
\bibfield{author}{\bibinfo{person}{QuillBot}.} \bibinfo{year}{2023}\natexlab{}.
\newblock \bibinfo{title}{Paraphraser AI tool}.
\newblock \bibinfo{howpublished}{\url{https://quillbot.com/}}.
\newblock
\newblock
\shownote{Accessed: 1 November 2023}.


\bibitem[Rauchfleisch and Kaiser(2020)]%
        {rauchfleisch2020false}
\bibfield{author}{\bibinfo{person}{Adrian Rauchfleisch} {and} \bibinfo{person}{Jonas Kaiser}.} \bibinfo{year}{2020}\natexlab{}.
\newblock \showarticletitle{The false positive problem of automatic bot detection in social science research}.
\newblock \bibinfo{journal}{\emph{PloS one}} \bibinfo{volume}{15}, \bibinfo{number}{10} (\bibinfo{year}{2020}), \bibinfo{pages}{e0241045}.
\newblock


\bibitem[Reimers and Gurevych(2019)]%
        {reimers-2019-sentence-bert}
\bibfield{author}{\bibinfo{person}{Nils Reimers} {and} \bibinfo{person}{Iryna Gurevych}.} \bibinfo{year}{2019}\natexlab{}.
\newblock \showarticletitle{Sentence-BERT: Sentence Embeddings using Siamese BERT-Networks}. In \bibinfo{booktitle}{\emph{Proceedings of the 2019 Conference on Empirical Methods in Natural Language Processing}}. \bibinfo{publisher}{Association for Computational Linguistics}.
\newblock
\urldef\tempurl%
\url{http://arxiv.org/abs/1908.10084}
\showURL{%
\tempurl}


\bibitem[Rovito et~al\mbox{.}(2022)]%
        {rovito2022evolutionary}
\bibfield{author}{\bibinfo{person}{Luigi Rovito}, \bibinfo{person}{Lorenzo Bonin}, \bibinfo{person}{Luca Manzoni}, {and} \bibinfo{person}{Andrea De~Lorenzo}.} \bibinfo{year}{2022}\natexlab{}.
\newblock \showarticletitle{An Evolutionary Computation Approach for Twitter Bot Detection}.
\newblock \bibinfo{journal}{\emph{Applied Sciences}} \bibinfo{volume}{12}, \bibinfo{number}{12} (\bibinfo{year}{2022}), \bibinfo{pages}{5915}.
\newblock


\bibitem[Sayyadiharikandeh et~al\mbox{.}(2020)]%
        {ensemblebotometer}
\bibfield{author}{\bibinfo{person}{Mohsen Sayyadiharikandeh}, \bibinfo{person}{Onur Varol}, \bibinfo{person}{Kai-Cheng Yang}, \bibinfo{person}{Alessandro Flammini}, {and} \bibinfo{person}{Filippo Menczer}.} \bibinfo{year}{2020}\natexlab{}.
\newblock \showarticletitle{Detection of Novel Social Bots by Ensembles of Specialized Classifiers}. In \bibinfo{booktitle}{\emph{Proceedings of the 29th ACM International Conference on Information \& Knowledge Management}} (Virtual Event, Ireland) \emph{(\bibinfo{series}{CIKM '20})}. \bibinfo{pages}{2725–2732}.
\newblock
\showISBNx{9781450368599}


\bibitem[Shao et~al\mbox{.}(2018)]%
        {shao2018spread}
\bibfield{author}{\bibinfo{person}{Chengcheng Shao}, \bibinfo{person}{Giovanni~Luca Ciampaglia}, \bibinfo{person}{Onur Varol}, \bibinfo{person}{Kai-Cheng Yang}, \bibinfo{person}{Alessandro Flammini}, {and} \bibinfo{person}{Filippo Menczer}.} \bibinfo{year}{2018}\natexlab{}.
\newblock \showarticletitle{The spread of low-credibility content by social bots}.
\newblock \bibinfo{journal}{\emph{Nature communications}} \bibinfo{volume}{9}, \bibinfo{number}{1} (\bibinfo{year}{2018}), \bibinfo{pages}{1--9}.
\newblock


\bibitem[Shu et~al\mbox{.}(2020)]%
        {shu2020fakenewsnet}
\bibfield{author}{\bibinfo{person}{Kai Shu}, \bibinfo{person}{Deepak Mahudeswaran}, \bibinfo{person}{Suhang Wang}, \bibinfo{person}{Dongwon Lee}, {and} \bibinfo{person}{Huan Liu}.} \bibinfo{year}{2020}\natexlab{}.
\newblock \showarticletitle{Fakenewsnet: A data repository with news content, social context, and spatiotemporal information for studying fake news on social media}.
\newblock \bibinfo{journal}{\emph{Big data}} \bibinfo{volume}{8}, \bibinfo{number}{3} (\bibinfo{year}{2020}), \bibinfo{pages}{171--188}.
\newblock


\bibitem[Shulman(2010)]%
        {SHULMAN20105}
\bibfield{author}{\bibinfo{person}{Amichai Shulman}.} \bibinfo{year}{2010}\natexlab{}.
\newblock \showarticletitle{The underground credentials market}.
\newblock \bibinfo{journal}{\emph{Computer Fraud \& Security}} \bibinfo{volume}{2010}, \bibinfo{number}{3} (\bibinfo{year}{2010}), \bibinfo{pages}{5--8}.
\newblock
\showISSN{1361-3723}
\urldef\tempurl%
\url{https://doi.org/10.1016/S1361-3723(10)70022-1}
\showDOI{\tempurl}


\bibitem[Somepalli et~al\mbox{.}(2021)]%
        {somepalli2021saint}
\bibfield{author}{\bibinfo{person}{Gowthami Somepalli}, \bibinfo{person}{Micah Goldblum}, \bibinfo{person}{Avi Schwarzschild}, \bibinfo{person}{C~Bayan Bruss}, {and} \bibinfo{person}{Tom Goldstein}.} \bibinfo{year}{2021}\natexlab{}.
\newblock \showarticletitle{Saint: Improved neural networks for tabular data via row attention and contrastive pre-training}.
\newblock \bibinfo{journal}{\emph{arXiv preprint arXiv:2106.01342}} (\bibinfo{year}{2021}).
\newblock


\bibitem[Tabassum et~al\mbox{.}(2023)]%
        {tabassum2023many}
\bibfield{author}{\bibinfo{person}{Fatima Tabassum}, \bibinfo{person}{Sameera Mubarak}, \bibinfo{person}{Lin Liu}, {and} \bibinfo{person}{Jia~Tina Du}.} \bibinfo{year}{2023}\natexlab{}.
\newblock \showarticletitle{How Many Features Do We Need to Identify Bots on Twitter?}. In \bibinfo{booktitle}{\emph{International Conference on Information}}. Springer, \bibinfo{pages}{312--327}.
\newblock


\bibitem[Tian et~al\mbox{.}(2020)]%
        {tian2020makes}
\bibfield{author}{\bibinfo{person}{Yonglong Tian}, \bibinfo{person}{Chen Sun}, \bibinfo{person}{Ben Poole}, \bibinfo{person}{Dilip Krishnan}, \bibinfo{person}{Cordelia Schmid}, {and} \bibinfo{person}{Phillip Isola}.} \bibinfo{year}{2020}\natexlab{}.
\newblock \showarticletitle{What makes for good views for contrastive learning?}
\newblock \bibinfo{journal}{\emph{Advances in neural information processing systems}}  \bibinfo{volume}{33} (\bibinfo{year}{2020}), \bibinfo{pages}{6827--6839}.
\newblock


\bibitem[Ucar et~al\mbox{.}(2021)]%
        {ucar2021subtab}
\bibfield{author}{\bibinfo{person}{Talip Ucar}, \bibinfo{person}{Ehsan Hajiramezanali}, {and} \bibinfo{person}{Lindsay Edwards}.} \bibinfo{year}{2021}\natexlab{}.
\newblock \showarticletitle{Subtab: Subsetting features of tabular data for self-supervised representation learning}.
\newblock \bibinfo{journal}{\emph{Advances in Neural Information Processing Systems}}  \bibinfo{volume}{34} (\bibinfo{year}{2021}), \bibinfo{pages}{18853--18865}.
\newblock


\bibitem[Van~Buuren and Groothuis-Oudshoorn(2011)]%
        {van2011mice}
\bibfield{author}{\bibinfo{person}{Stef Van~Buuren} {and} \bibinfo{person}{Karin Groothuis-Oudshoorn}.} \bibinfo{year}{2011}\natexlab{}.
\newblock \showarticletitle{mice: Multivariate imputation by chained equations in R}.
\newblock \bibinfo{journal}{\emph{Journal of statistical software}}  \bibinfo{volume}{45} (\bibinfo{year}{2011}), \bibinfo{pages}{1--67}.
\newblock


\bibitem[Varol et~al\mbox{.}(2017)]%
        {varol2017online}
\bibfield{author}{\bibinfo{person}{Onur Varol}, \bibinfo{person}{Emilio Ferrara}, \bibinfo{person}{Clayton Davis}, \bibinfo{person}{Filippo Menczer}, {and} \bibinfo{person}{Alessandro Flammini}.} \bibinfo{year}{2017}\natexlab{}.
\newblock \showarticletitle{Online human-bot interactions: Detection, estimation, and characterization}. In \bibinfo{booktitle}{\emph{Proceedings of the international AAAI conference on web and social media}}, Vol.~\bibinfo{volume}{11}. \bibinfo{pages}{280--289}.
\newblock


\bibitem[Vaughan-Nichols(2023)]%
        {darkweb}
\bibfield{author}{\bibinfo{person}{Steven Vaughan-Nichols}.} \bibinfo{year}{2023}\natexlab{}.
\newblock \bibinfo{title}{Hacked! My Twitter user data is out on the dark web -- now what?}
\newblock \bibinfo{howpublished}{\url{https://www.zdnet.com/article/hacked-my-twitter-user-data-is-out-on-the-dark-web-now-what/}}.
\newblock
\newblock
\shownote{Accessed: 18 September 2023}.


\bibitem[Vosoughi et~al\mbox{.}(2018)]%
        {vosoughi2018spread}
\bibfield{author}{\bibinfo{person}{Soroush Vosoughi}, \bibinfo{person}{Deb Roy}, {and} \bibinfo{person}{Sinan Aral}.} \bibinfo{year}{2018}\natexlab{}.
\newblock \showarticletitle{The spread of true and false news online}.
\newblock \bibinfo{journal}{\emph{science}} \bibinfo{volume}{359}, \bibinfo{number}{6380} (\bibinfo{year}{2018}), \bibinfo{pages}{1146--1151}.
\newblock


\bibitem[Wright and Anise(2018)]%
        {wright2018don}
\bibfield{author}{\bibinfo{person}{Jordan Wright} {and} \bibinfo{person}{Olabode Anise}.} \bibinfo{year}{2018}\natexlab{}.
\newblock \showarticletitle{Don’t@ me: Hunting twitter bots at scale}.
\newblock \bibinfo{journal}{\emph{Blackhat USA}} (\bibinfo{year}{2018}).
\newblock


\bibitem[write(2018)]%
        {howmanywords}
\bibfield{author}{\bibinfo{person}{Writers write}.} \bibinfo{year}{2018}\natexlab{}.
\newblock \bibinfo{title}{How Many Words is That?}
\newblock \bibinfo{howpublished}{\url{https://tinyurl.com/how-many-words-in-tweets/}}.
\newblock
\newblock
\shownote{Accessed: 18 September 2023}.


\bibitem[Wu et~al\mbox{.}(2020)]%
        {wu2020clear}
\bibfield{author}{\bibinfo{person}{Zhuofeng Wu}, \bibinfo{person}{Sinong Wang}, \bibinfo{person}{Jiatao Gu}, \bibinfo{person}{Madian Khabsa}, \bibinfo{person}{Fei Sun}, {and} \bibinfo{person}{Hao Ma}.} \bibinfo{year}{2020}\natexlab{}.
\newblock \showarticletitle{Clear: Contrastive learning for sentence representation}.
\newblock \bibinfo{journal}{\emph{arXiv preprint arXiv:2012.15466}} (\bibinfo{year}{2020}).
\newblock


\bibitem[Yang et~al\mbox{.}(2022)]%
        {botometer}
\bibfield{author}{\bibinfo{person}{Kai-Cheng Yang}, \bibinfo{person}{Emilio Ferrara}, {and} \bibinfo{person}{Filippo Menczer}.} \bibinfo{year}{2022}\natexlab{}.
\newblock \showarticletitle{Botometer 101: social bot practicum for computational social scientists}.
\newblock \bibinfo{journal}{\emph{Journal of Computational Social Science}} \bibinfo{volume}{5}, \bibinfo{number}{2} (\bibinfo{date}{01 Nov} \bibinfo{year}{2022}).
\newblock


\bibitem[Yang and Menczer(2023)]%
        {yang2023anatomy}
\bibfield{author}{\bibinfo{person}{Kai-Cheng Yang} {and} \bibinfo{person}{Filippo Menczer}.} \bibinfo{year}{2023}\natexlab{}.
\newblock \showarticletitle{Anatomy of an AI-powered malicious social botnet}.
\newblock \bibinfo{journal}{\emph{arXiv preprint arXiv:2307.16336}} (\bibinfo{year}{2023}).
\newblock


\bibitem[Yang et~al\mbox{.}(2020)]%
        {Yang_Varol_Hui_Menczer_2020}
\bibfield{author}{\bibinfo{person}{Kai-Cheng Yang}, \bibinfo{person}{Onur Varol}, \bibinfo{person}{Pik-Mai Hui}, {and} \bibinfo{person}{Filippo Menczer}.} \bibinfo{year}{2020}\natexlab{}.
\newblock \showarticletitle{Scalable and Generalizable Social Bot Detection through Data Selection}.
\newblock \bibinfo{journal}{\emph{Proceedings of the AAAI Conference on Artificial Intelligence}} \bibinfo{volume}{34}, \bibinfo{number}{01} (\bibinfo{date}{Apr.} \bibinfo{year}{2020}), \bibinfo{pages}{1096--1103}.
\newblock
\urldef\tempurl%
\url{https://doi.org/10.1609/aaai.v34i01.5460}
\showDOI{\tempurl}


\bibitem[Yin et~al\mbox{.}(2020)]%
        {yin2020tabert}
\bibfield{author}{\bibinfo{person}{Pengcheng Yin}, \bibinfo{person}{Graham Neubig}, \bibinfo{person}{Wen-tau Yih}, {and} \bibinfo{person}{Sebastian Riedel}.} \bibinfo{year}{2020}\natexlab{}.
\newblock \showarticletitle{TaBERT: Pretraining for joint understanding of textual and tabular data}.
\newblock \bibinfo{journal}{\emph{arXiv preprint arXiv:2005.08314}} (\bibinfo{year}{2020}).
\newblock


\bibitem[Yoon et~al\mbox{.}(2020)]%
        {yoon2020vime}
\bibfield{author}{\bibinfo{person}{Jinsung Yoon}, \bibinfo{person}{Yao Zhang}, \bibinfo{person}{James Jordon}, {and} \bibinfo{person}{Mihaela van~der Schaar}.} \bibinfo{year}{2020}\natexlab{}.
\newblock \showarticletitle{Vime: Extending the success of self-and semi-supervised learning to tabular domain}.
\newblock \bibinfo{journal}{\emph{Advances in Neural Information Processing Systems}}  \bibinfo{volume}{33} (\bibinfo{year}{2020}), \bibinfo{pages}{11033--11043}.
\newblock


\bibitem[Yun et~al\mbox{.}(2019)]%
        {yun2019cutmix}
\bibfield{author}{\bibinfo{person}{Sangdoo Yun}, \bibinfo{person}{Dongyoon Han}, \bibinfo{person}{Seong~Joon Oh}, \bibinfo{person}{Sanghyuk Chun}, \bibinfo{person}{Junsuk Choe}, {and} \bibinfo{person}{Youngjoon Yoo}.} \bibinfo{year}{2019}\natexlab{}.
\newblock \showarticletitle{Cutmix: Regularization strategy to train strong classifiers with localizable features}. In \bibinfo{booktitle}{\emph{Proceedings of the IEEE/CVF international conference on computer vision}}. \bibinfo{pages}{6023--6032}.
\newblock


\bibitem[Zhai et~al\mbox{.}(2019)]%
        {zhai2019s4l}
\bibfield{author}{\bibinfo{person}{Xiaohua Zhai}, \bibinfo{person}{Avital Oliver}, \bibinfo{person}{Alexander Kolesnikov}, {and} \bibinfo{person}{Lucas Beyer}.} \bibinfo{year}{2019}\natexlab{}.
\newblock \showarticletitle{S4l: Self-supervised semi-supervised learning}. In \bibinfo{booktitle}{\emph{Proceedings of the IEEE/CVF international conference on computer vision}}. \bibinfo{pages}{1476--1485}.
\newblock


\bibitem[Zhang et~al\mbox{.}(2017)]%
        {zhang2017mixup}
\bibfield{author}{\bibinfo{person}{Hongyi Zhang}, \bibinfo{person}{Moustapha Cisse}, \bibinfo{person}{Yann~N Dauphin}, {and} \bibinfo{person}{David Lopez-Paz}.} \bibinfo{year}{2017}\natexlab{}.
\newblock \showarticletitle{mixup: Beyond empirical risk minimization}.
\newblock \bibinfo{journal}{\emph{arXiv preprint arXiv:1710.09412}} (\bibinfo{year}{2017}).
\newblock


\bibitem[Zhang et~al\mbox{.}(2022)]%
        {zhang2022contrastive}
\bibfield{author}{\bibinfo{person}{Rui Zhang}, \bibinfo{person}{Yangfeng Ji}, \bibinfo{person}{Yue Zhang}, {and} \bibinfo{person}{Rebecca~J Passonneau}.} \bibinfo{year}{2022}\natexlab{}.
\newblock \showarticletitle{Contrastive data and learning for natural language processing}. In \bibinfo{booktitle}{\emph{Proceedings of the 2022 Conference of the North American Chapter of the Association for Computational Linguistics: Human Language Technologies: Tutorial Abstracts}}. \bibinfo{pages}{39--47}.
\newblock


\bibitem[Zhou et~al\mbox{.}(2023)]%
        {zhou2023detecting}
\bibfield{author}{\bibinfo{person}{Ming Zhou}, \bibinfo{person}{Dan Zhang}, \bibinfo{person}{Yuandong Wang}, \bibinfo{person}{Yangli-Ao Geng}, {and} \bibinfo{person}{Jie Tang}.} \bibinfo{year}{2023}\natexlab{}.
\newblock \showarticletitle{Detecting Social Bot on the Fly using Contrastive Learning}. In \bibinfo{booktitle}{\emph{Proceedings of the 32nd ACM International Conference on Information and Knowledge Management}}. \bibinfo{pages}{4995--5001}.
\newblock


\end{thebibliography}

\appendix

\section{Visualization of Social Bot Datasets}
\label{appendix:visualization}
From Figure~\ref{fig:problem}, we can see the last three datasets have separability problems for the classifier due to low homogeneity within clusters or similar features for both classes (bot and human). The most intuitive way to inspect the different datasets is to visualize them in feature space. We take this result of inspecting different datasets from [7], where the authors projected each dataset into 2-D plane for visualization. To quantify the separation of bots and human accounts in each dataset, the authors applied kNN classifier in the original feature space. With the labels obtained from kNN and the ground truth, they were able to calculate the homogeneity score for each dataset. From Figure~\ref{fig:problem}, five of 11 datasets demonstrate a clearly clustered structure, suggesting bots and humans are easily separable using feature-based. The rest of the datasets have clusters that are not as easily separable. The same figure shows that accounts in the Gilani dataset had lower homogeneity scores than accounts in the Varol dataset, which refers to the KNN cluster homogeneity score achieved in ~\cite{Yang_Varol_Hui_Menczer_2020}.
\begin{figure*}[!htb]
  \centering
  \includegraphics[width=\linewidth]{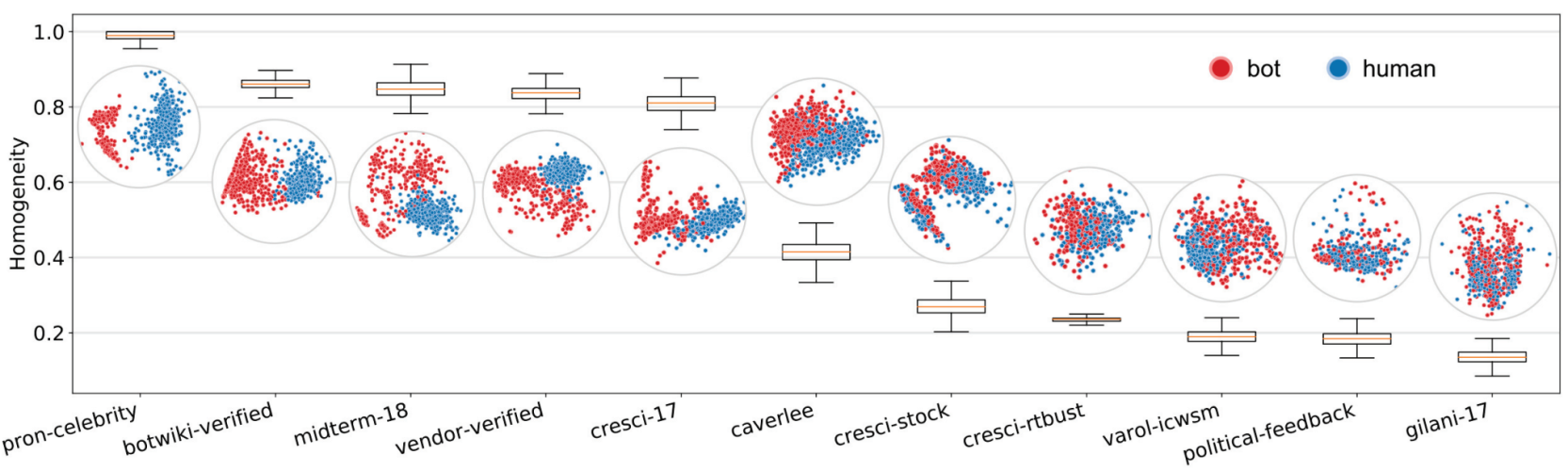}
  \caption{Visualization of human and bot accounts in different training datasets, Figure Source: ~\cite{Yang_Varol_Hui_Menczer_2020}. The last three datasets (especially Varol and Gilani) show the minimum homogeneity and separability problems between the bot and human classes.}
  \label{fig:problem}
\end{figure*}

\section{Features Used in our work}
\label{appendix:features_used}
Table~\ref{tab:features_elaborated} shows the different features used in our work. The table particularly displays different feature category types and the data collection of tier level it falls into for feature engineering. 
\begin{table*}[!htb]
   \caption{Features used in our work for Twitter Bot Detection.} 
\label{tab:features_elaborated}
\scalebox{0.76}{
  \begin{tabular}{c c l c}
   \toprule
  \textbf{Feature Type}& \textbf{Feature No.}& \textbf{Feature Name}& \textbf{Data Collection Tier}\\
   \hline
      \multirow{33}{*}{User Metadata} & 1 & followers\_count & \multirow{33}{*}{Tier 0} \\ 
   & 2 & friends\_count & \\ 
   & 3 & listed\_count & \\ 
   & 4 & verified & \\ 
   & 5 & user\_age & \\ 
   & 6 & follower\_growth\_rate & \\ 
   & 7 & friends\_growth\_rate & \\ 
   & 8 & listed\_growth\_rate & \\
   & 9 & followers\_friend\_ratio & \\ 
   & 10 & name\_length & \\
   & 11 & username\_length & \\ 
   & 12 & description\_length & \\ 
   & 13 & num\_digits\_in\_name & \\ 
   & 14 & num\_digits\_in\_username & \\ 
   & 15 & names\_ratio & \\
   & 16 & name\_freq & \\
   & 17 & name\_entropy & \\
   & 18 & username\_entropy & \\ 
   & 19 & description\_entropy & \\ 
   & 20 & description\_sentiment & \\ 
   & 21 & names\_sim & \\ 
   & 22 & url\_in\_description & \\ 
   & 23 & bot\_in\_names & \\ 
   & 24 & hashtag\_in\_description & \\ 
   & 25 & hashtag\_in\_name & \\ 
   & 26 & numbers\_in\_description & \\ 
   & 27 & numbers\_in\_name & \\ 
   & 28 & numbers\_in\_username & \\ 
   & 29 & emojis\_in\_description & \\ 
   & 30 & emojis\_in\_name & \\ 
   & 31 & favourites\_count & \\ 
   & 32 & status\_count & \\ 
   & 33 & default\_profile & \\ \hdashline
   User Tweets& 1 & average BERT embedding of user's 200 tweets textual content & Tier 1\\
\hdashline
   \multirow{29}{*}{Tweet Metadata} & 1 & mean\_no\_emoticons & \multirow{29}{*}{Tier 1} \\ 
   & 2 & mean\_no\_urls\_per\_tweet & \\ 
   & 3 & mean\_no\_media\_per\_tweet & \\ 
   & 4 & mean\_no\_words & \\ 
   & 5 & no\_languages & \\
   & 6 & mean\_no\_hashtags & \\ 
   & 7 & mean\_number\_of\_positive\_emoticons\_per\_tweet & \\ 
   & 8 & mean\_number\_of\_negative\_emoticons\_per\_tweet & \\ 
   & 9 & mean\_number\_of\_neutral\_emoticons\_per\_tweet & \\ 
   & 10 & mean\_tweet\_sentiment & \\ 
   & 11 & mean\_positive\_valence\_score\_per\_tweet & \\ 
   & 12 & mean\_negative\_valence\_score\_per\_tweet & \\ 
   & 13 & mean\_neutral\_valence\_score\_per\_tweet & \\ 
   & 14 & positive\_valence\_score\_of\_aggregated\_tweets & \\ 
   & 15 & negative\_valence\_score\_of\_aggregated\_tweets & \\ 
   & 16 & neutral\_valence\_score\_of\_aggregated\_tweets & \\ 
   & 17 & mean\_positive\_and\_negative\_score\_ratio\_per\_tweet & \\ 
   & 18 & mean\_emoticons\_entropy\_per\_tweet \\ 
   & 19 & mean\_emoticons\_entropy\_of\_aggregated\_tweets & \\ 
   & 20 & mean\_negative\_emoticons\_entropy\_of\_aggregated\_tweets & \\ 
   & 21 & mean\_positive\_emoticons\_entropy\_of\_aggregated\_tweets & \\ 
   & 22 & mean\_neutral\_emoticons\_entropy\_of\_aggregated\_tweets & \\ 
   & 23 & mean\_positive\_emoticons\_entropy\_per\_tweet & \\ 
   & 24 & mean\_negative\_emoticons\_entropy\_per\_tweet & \\ 
   & 25 & mean\_neutral\_emoticons\_entropy\_per\_tweet & \\ 
   & 26 & mean\_favourites\_per\_tweet & \\ 
   & 27 & mean\_retweets\_per\_tweet & \\ 
   & 28 & no\_retweet\_tweets & \\ 
   & 29 & retweet\_as\_tweet\_rate & \\ 
   \hdashline

   \multirow{7}{*}{Tweet Temporal} & 1 & time\_between\_tweets & \multirow{7}{*}{Tier 1} \\ 
   & 2 & tweet\_frequency & \\ 
   & 3 & min\_tweets\_per\_hour & \\ 
   & 4 & min\_tweets\_per\_day & \\ 
   & 5 & max\_tweets\_per\_hour & \\ 
   & 6 & max\_tweets\_per\_day & \\ 
   & 7 & max\_occurence\_of\_same\_gap & \\ 
   \bottomrule
  \end{tabular}
  }
\end{table*}


\section{Experimental Setup}
\label{appendix:experimentsetup}
Our BotSSCL is a variant of \textit{SCARF}~\cite{bahri2021scarf} implemented in PyTorch version 1.12 with CUDA 10.7. We conducted our performance assessments using a server consisting of 2x Intel Xeon Silver 4208 processors with 128 GB of RAM, all running Ubuntu 20.04.01 LTS. The server has 4 NVIDIA Tesla T4 GPUs, each with 16GB of RAM, yet we specifically utilized a single GPU throughout the experiments. The model is trained using Adam optimizer with a default temperate of 1, no dropout, and a learning rate 0.001.

\section{All Evaluation Results}
\label{appendix:performance}
Table~\ref{tab:all_evalutation_varol} and \ref{tab:all_evalutation_gilani} displays all performed measurements and evaluations. The table specifies which dataset, augmentation technique, input, output dimension, and \textit{D} value have been used for the training and test set. The table contains the Precision, Recall, F1-score, and Accuracy achieved in percentages. As it is seen from Table~\ref{tab:all_evalutation_varol}, 16 input dimensions and  64 output dimensions are the best configurations (can also be written as 16$\times$64) for the Varol dataset achieving 80\%. On the other hand, the Gilani dataset works best with 64 as the input dimension and 64 as the output embedding dimension (64$\times$64), as depicted in Table~\ref{tab:all_evalutation_gilani}. Both results are optimal with Augmentation 1 (feature corruption) with 0.6 as the corruption rate. It is worth noting that the Augmentation 2 result for input and \textit{D} feature dimension of more than 32 dimensions is null as the MICE method is computationally expensive for larger dimensions. Therefore, it is not presented in the table.
Thus, only result till 32$\times$32 is achieved on both datasets on augmentation 2. These results are used in Section~\ref{subsec:performance} and Ablation Test 1.

\begin{table*}[!htb]
   \caption{All evaluation cases. This table displays all performed measurements in (\%) across different augmentation techniques on the Varol Dataset. \textit{D} refers to the input feature dimension, Dx4 is the Twitter user representation dimension, and Output refers to the output embedding dimension.} 
\label{tab:all_evalutation_varol}
\scalebox{0.90}{
  \begin{tabular}{c c c c c c c c c c}
   \toprule
  \textbf{Test ID (DxOutput)}&\textbf{Dataset}& \textbf{Augmentation}& \textbf{\textit{D} Input Feature} & \textbf{(Dx4)} & \textbf{Output} & \textbf{Precision} & \textbf{Recall} &\textbf{F1} & \textbf{Accuracy}\\
   \hline
   16$\times$16 & Varol & 1 & 16 & 64 & 16 & 75 & 76 & 75 & 78 \\
   16$\times$32 & Varol & 1 & 16 & 64 & 32 & 75 & 76 & 76 & 78 \\
   \textbf{16$\times$64} & \textbf{Varol} & \textbf{1} & \textbf{16} & \textbf{64} & \textbf{64} &  \textbf{80}& \textbf{81} & \textbf{80} & \textbf{82} \\
    16$\times$128 & Varol & 1 & 16 & 64 & 128 & 78 & 80 & 79 & 81 \\ \hdashline
    32$\times$16 & Varol & 1 & 32 & 128 & 16 & 75 & 76 & 75 & 78 \\
    32$\times$32 & Varol & 1 & 32 & 128 & 32 & 77 & 78 & 77 & 80 \\
    32$\times$64 & Varol & 1 & 32 & 128 & 64 & 76 & 78 & 77 & 79 \\
    32$\times$128 & Varol & 1 & 32 & 128 & 128 & 78 & 80 & 78 & 80 \\ \hdashline
    64$\times$16 & Varol & 1 & 64 & 256 & 16 & 76 & 78 & 77 & 79 \\
    64$\times$32 & Varol & 1 & 64 & 256 & 32 & 76 & 78 & 77 & 79 \\
    64$\times$64 & Varol & 1 & 64 & 256 & 64 & 76 & 78 & 77 & 79 \\
    64$\times$128 & Varol & 1 & 64 & 256 & 128 & 78 & 79 & 78 & 80 \\ \hdashline
    128$\times$16 & Varol & 1 & 128 & 512 & 16 & 76 & 78 & 77 & 79 \\
    128$\times$32 & Varol & 1 & 128 & 512 & 32 & 76 & 77 & 77 & 80 \\
    128$\times$64 & Varol & 1 & 128 & 512 & 64 & 78 & 80 & 79 & 81 \\
    128$\times$128 & Varol & 1 & 128 & 512 & 128 & 77 & 79 & 77 & 80 \\ \hdashline
    16$\times$16 & Varol & 2 & 16 & 64 & 16 & 71 & 73 & 72 & 75 \\
    16$\times$32 & Varol & 2 & 16 & 64 & 32 & 71 & 73 & 72 & 75 \\
    32$\times$16 & Varol & 2 & 32 & 128 & 16 & 71 & 72 & 71 & 75 \\
    32$\times$32 & Varol & 2 & 32 & 128 & 32 & 73 & 75 & 73 & 77 \\ \hdashline
    16$\times$16 & Varol & 3 & 16 & 64 & 16 & 72 & 72 & 72 & 76 \\
    16$\times$32 & Varol & 3 & 16 & 64 & 32 & 68 & 69 & 69 & 72 \\
    16$\times$64 & Varol & 3 & 16 & 64 & 64 & 76 & 77 & 76 & 79 \\
    16$\times$128 & Varol & 3 & 16 & 64 & 128 & 75 & 76 & 76 & 79 \\ \hdashline
    32$\times$16 & Varol & 3 & 32 & 128 & 16 & 63 & 64 & 63 & 67 \\
    32$\times$32 & Varol & 3 & 32 & 128 & 32 & 72 & 74 & 72 & 75 \\
    32$\times$64 & Varol & 3 & 32 & 128 & 64 & 71 & 73 & 72 & 75 \\
    32$\times$128 & Varol & 3 & 32 & 128 & 128 & 76 & 77 & 76 & 79 \\ \hdashline
    64$\times$16 & Varol & 3 & 64 & 256 & 16 & 67 & 68 & 67 & 70 \\
    64$\times$32 & Varol & 3 & 64 & 256 & 32 & 77 & 77 & 77 & 80 \\
    64$\times$64 & Varol & 3 & 64 & 256 & 64 & 74 & 73 & 73 & 77 \\
    64$\times$128 & Varol & 3 & 64 & 256 & 128 & 74 & 76 & 75 & 77 \\ \hdashline
    128$\times$16 & Varol & 3 & 128 & 512 & 16 & 70 & 71 & 71 & 74 \\
    128$\times$32 & Varol & 3 & 128 & 512 & 32 & 73 & 75 & 74 & 76 \\
    128$\times$64 & Varol & 3 & 128 & 512 & 64 & 76 & 77 & 76 & 79 \\
    128$\times$128 & Varol & 3 & 128 & 512 & 128 & 75 & 76 & 75 & 78 \\
   \bottomrule
  \end{tabular}}
\end{table*}

\begin{table*}[!htb]
   \caption{All evaluation cases. This table displays all performed measurements in (\%) across different augmentation techniques on the Gilani Dataset. \textit{D} refers to the input feature dimension, D$\times$4 is the Twitter user representation dimension, and Output refers to the output embedding dimension.} 
\label{tab:all_evalutation_gilani}
\scalebox{0.90}{
  \begin{tabular}{c c c c c c c c c c}
   \toprule
  \textbf{Test ID (DxOutput)}&\textbf{Dataset}& \textbf{Augmentation}& \textbf{\textit{D} Input Feature} & \textbf{(D$\times$4)} & \textbf{Output} & \textbf{Precision} & \textbf{Recall} &\textbf{F1} & \textbf{Accuracy}\\
   \hline
    16$\times$16 & Gilani & 1 & 16 & 64 & 16 & 76 & 76 & 76 & 77 \\
    16$\times$32 & Gilani & 1 & 16 & 64 & 32 & 76 & 76 & 76 & 77 \\
    16$\times$64 & Gilani & 1 & 16 & 64 & 64 & 78 & 78 & 78 & 78 \\
    16$\times$128 & Gilani & 1 & 16 & 64 & 128 & 78 & 78 & 78 & 78 \\ \hdashline
    32$\times$16 & Gilani & 1 & 32 & 128 & 16 & 77 & 76 & 76 & 77 \\
    32$\times$32 & Gilani & 1 & 32 & 128 & 32 & 75 & 75 & 75 & 76 \\
    32$\times$64 & Gilani & 1 & 32 & 128 & 64 & 77 & 76 & 76 & 77 \\
    32$\times$128 & Gilani & 1 & 32 & 128 & 128 & 78 & 78 & 78 & 78 \\ \hdashline
    64$\times$16 & Gilani & 1 & 64 & 256 & 16 & 77 & 77 & 77 & 77 \\
    64$\times$32 & Gilani & 1 & 64 & 256 & 32 & 77 & 77 & 77 & 78 \\
   \textbf{64$\times$64} & \textbf{Gilani} & \textbf{1} & \textbf{64} & \textbf{256} & \textbf{64} & \textbf{79} & \textbf{79} & \textbf{79} & \textbf{80} \\ 
    64$\times$128 & Gilani & 1 & 64 & 256 & 128 & 77 & 77 & 77 & 78 \\ \hdashline
    128$\times$16 & Gilani & 1 & 128 & 512 & 16 & 78 & 77 & 77 & 78 \\
    128$\times$32 & Gilani & 1 & 128 & 512 & 32 & 76 & 76 & 76 & 77 \\
    128$\times$64 & Gilani & 1 & 128 & 512 & 64 & 79 & 78 & 78 & 79 \\
    128$\times$128 & Gilani & 1 & 128 & 512 & 128 & 77 & 77 & 77 & 77 \\ \hdashline
    16$\times$16 & Gilani & 2 & 16 & 64 & 16 & 70 & 69 & 70 & 71 \\
    16$\times$32 & Gilani & 2 & 16 & 64 & 32 & 70 & 70 & 70 & 71 \\ 
    32$\times$16 & Gilani & 2 & 32 & 128 & 16 & 68 & 67 & 68 & 69 \\
    32$\times$32 & Gilani & 2 & 32 & 128 & 32 & 72 & 71 & 71 & 73 \\ \hdashline
    16$\times$16 & Gilani & 3 & 16 & 64 & 16 & 69 & 69 & 69 & 69 \\
    16$\times$32 & Gilani & 3 & 16 & 64 & 32 & 71 & 70 & 70 & 71 \\
    16$\times$64 & Gilani & 3 & 16 & 64 & 64 & 71 & 71 & 71 & 71 \\
    16$\times$128 & Gilani & 3 & 16 & 64 & 128 & 73 & 73 & 73 & 74 \\ \hdashline
    32$\times$16 & Gilani & 3 & 32 & 128 & 16 & 70 & 70 & 70 & 71 \\
    32$\times$32 & Gilani & 3 & 32 & 128 & 32 & 72 & 72 & 72 & 72 \\
    32$\times$64 & Gilani & 3 & 32 & 128 & 64 & 74 & 75 & 74 & 75 \\
    32$\times$128 & Gilani & 3 & 32 & 128 & 128 & 74 & 74 & 74 & 75 \\ \hdashline
    64$\times$16 & Gilani & 3 & 64 & 256 & 16 & 75 & 73 & 74 & 75 \\
    64$\times$32 & Gilani & 3 & 64 & 256 & 32 & 68 & 69 & 68 & 69 \\
    64$\times$64 & Gilani & 3 & 64 & 256 & 64 & 74 & 73 & 73 & 75 \\
    64$\times$128 & Gilani & 3 & 64 & 256 & 128 & 77 & 77 & 77 & 78 \\ \hdashline
    128$\times$16 & Gilani & 3 & 128 & 512 & 16 & 69 & 68 & 69 & 70 \\
    128$\times$32 & Gilani & 3 & 128 & 512 & 32 & 74 & 73 & 74 & 74 \\
    128$\times$64 & Gilani & 3 & 128 & 512 & 64 & 75 & 75 & 75 & 76 \\
    128$\times$128 & Gilani & 3 & 128 & 512 & 128 & 75 & 75 & 75 & 76 \\
   \bottomrule
  \end{tabular}}
\end{table*}

\section{Features used in Adversarial Attack}
\label{appendix:adversarial}
This appendix discusses the features used in adversarial perturbations to inputs (cf. \S\ref{subsec:adversarial}) and the reasoning behind their selection.  
\begin{itemize}[wide, labelwidth=!, labelindent=0pt]
\item In the \textit{user metadata} feature, an attacker can feasibly alter two features, i.e., number of followers and number of followings. As previously defined, attackers can make all purchased accounts follow the master account. It means there is a possibility of values from 0 (no followers) to 1000 (all fake followers). 

\item In the \textit{tweet metadata} feature, an adversary can modify three features. First, they can change the \textit{mean number of words} in a tweet. Considering a 5.1 average number of characters per word, an adversary can use either one word or 55 words in a tweet (as Twitter allows 280 characters at most)~\cite{howmanywords}. Second, the adversary can control the \textit{mean number of favorites (likes)} per tweet. There is a possibility of values from 0 (no likes) to 1000 (all fake likes). Lastly, as we know, it is easier for a bot account owner to retweet instead of producing original content~\cite{akhtar2023false}. It should be noted that Twitter also considers retweets as tweets, and there is a 2400 Twitter tweet limit per day~\cite{twitterratelimit}. However, in our case, we limit the value to 200 instead of 2400 as 200 relates to the number of tweets collected per account by our team during data collection from Twitter. In our threat model, the adversary can control the posting of the \textit{number of retweets}, thereby perturbing the ratio of retweets and tweets.

\item In the \textit{tweet temporal} feature, the adversary can modify two attributes, i.e., the maximum number of tweets per hour and day. Twitter limits the maximum number of tweets to 100 per hour and 2400 per day~\cite{tweetperhour,twitterratelimit}. However, in our case, the adversary can post a max. of 200 tweets daily and can post up to 100 tweets per hour. 

\item Finally, an adversary can perturb a combination of all the above features, thereby manipulating seven prominent features in total. 
\end{itemize}

\section{Choice for Adversarial Perturbations Initial Samples}
\label{appendix:adversarial_choice}
{The \textit{URET}~\cite{eykholt2023uret} allows an argument to set the number of samples to input to the model for adversarial perturbation. Practically, the number of the sample can be the whole dataset. However, for the adversarial evaluation of BotSSCL, we only provided part of the dataset (200 samples). We intended to show that it is computationally infeasible for an adversary to create many adversarial accounts.}

We opted for 200 as the appropriate number after testing with varying sample input sizes. Table~\ref{tab:adversarial_choice} shows that the time taken to generate adversarial samples increases as we increase the number of input samples (from 200 to 300). Both user and tweet metadata have features with an extensive range of possible values that URET chooses to modify; this is why we can see the increase in the time taken to create adversarial inputs. Therefore, choosing 200 due to the constraint factor is an appropriate value for adversarial robustness evaluation.

\begin{table*}[!h]
  \centering
   \caption{Evaluation of Adversarial Robustness of BotSSCL in terms of \textit{success rate},
   number of \textit{adversarial samples} generated,
   and \textit{time taken} to brute force complete search space out of 200 and 300 initial samples.}
   \vspace{-0.2cm}
\label{tab:adversarial_choice}
\scalebox{0.9}{
  \begin{tabular}{l c c l : c c l}
   \toprule
   \multirow{2}{*}{\textbf{Adversarial Manipulation}}&\multicolumn{3}{c}{\textbf{BotSSCL (out of 200 sample)}}&\multicolumn{3}{c}{\textbf{BotSSCL (out of 300 sample)}}\\
   \cline{2-7}
  &\textbf{Success Rate} & \textbf{Samples} &\textbf{Time Taken}&\textbf{Success Rate} & \textbf{Samples} &\textbf{Time Taken}\\
   \hline
   1) User Metadata Feature  & 0.5\% & 1 & $\approx 10$ Hours & 0 \% & 0 & $\approx$ 18 Hours \\
   2) Tweet Metadata Feature & 2.5 \% & 5 & $\approx 9$ Hours & 1.0 \%& 3 & $\approx$ 12.5 hours \\
   3) Tweet Temporal Feature  & 12.5 \% & 25 & $\approx 3$ Hours & 10.6 \% & 32 & $\approx$ 3 Hours \\ \hdashline
   4) All Above Three Feature  & 4.0 \% & 8 & $\approx 19.5$ Hours & 4.0 \% & 12 & $\approx$ 20 Hours \\
   \bottomrule
    \end{tabular}
  }
  \end{table*}

\section{Textual Feature Perturbations using Generative AI (ChatGPT)}
\label{appendix:adversarial_gpt}
We detail the adversary's capability to modify the textual content of tweets as below:

\begin{enumerate}[wide, labelwidth=!, labelindent=0pt]
    \item The adversary can pass \{\textit{input queries}\} for new generation/paraphrasing/re-written and receive \{\textit{output tokens}\} from generative AI such as ChatGPT~\cite{chatgpt}.
    \item The adversary can manipulate their tweet individually one by one and test whether the model flips the prediction.
    \item Receiving modified/paraphrased/re-written tweets from ChatGPT also changes the tweet metadata features, representing more accurate perturbations.
    \item We here show only 20 samples of bots with a minimum of 150 tweets (with a maximum of 200 collected tweets) posted in their timeline as ChatGPT has a constraint for every input/output token involved per query, which we refer to as a prompt budget the adversary is ready to spend.
\end{enumerate}
Our results show that out of 20 accounts, only one flipped the output after 193 prompts. Our analysis shows that textual adversarial attack success is at most 5\% as shown in Table~\ref{tab:adversarial_gpt}. 

\begin{table}[!h]
  \centering
   \caption{Evaluation of Adversarial Robustness of BotSSCL in terms of \textit{success rate},
   number of \textit{adversarial samples} generated,
   and \textit{time taken} to brute force complete search space.}
   \vspace{-0.45cm}
\label{tab:adversarial_gpt}
\scalebox{0.835}{
  \begin{tabular}{l c c l}
   \toprule
   \multirow{2}{*}{\textbf{Adversarial Manipulation}}&\multicolumn{3}{c}{\textbf{BotSSCL (out of 20 sample)}}\\
   \cline{2-4}
  &\textbf{Success Rate} & \textbf{Samples} &\textbf{Time Taken}\\
   \hline
   1) Textual Content Feature & 5.0\% & 1 & $\approx$ 5 Hours \\
   \bottomrule
  \end{tabular}
  }
  \end{table}

\textbf{Why did we choose ChatGPT?:} We find that TextAttack~\cite{morris2020textattack}, URET~\cite{eykholt2023uret}, and text spinners such as Quillbot~\cite{quillbot} already exist in addition to generative AI tools. We chose not to use TextAttack as it offers limited transformation of input queries via word substitution, such as substituting with the word `banana'. In addition, the URET model can aid in text manipulation but is built upon the existing libraries such as TextAttack, thus only offering new string generation based on substitution, addition, and deletion of characters. Finally, text spinners can generate paraphrases or unique output but do not always meet the 280-character Twitter tweet's requirement. Due to this, we opted for ChatGPT as the appropriate generative AI tool for the adversary.

\end{document}